\documentclass[conference,compsoc]{IEEEtran}
\IEEEoverridecommandlockouts
\usepackage[most]{tcolorbox}
\usepackage{booktabs} 
\usepackage{multirow} 
\usepackage{amssymb}
\usepackage{pifont}
\usepackage{multirow}
\usepackage{amsmath}
\usepackage{soul}
\usepackage{colortbl}
\usepackage{arydshln}
\usepackage{wasysym}
\usepackage{adjustbox}
\PassOptionsToPackage{table}{xcolor}
\usepackage[table]{xcolor}
\usepackage[T1]{fontenc}

\usepackage{epsfig,amsmath,amsfonts,epsfig,multirow,makecell,caption,soul,csquotes,color,wrapfig,subcaption,mathtools,bm,spverbatim,booktabs,tcolorbox,diagbox}
\usepackage[e]{esvect}

\makeatletter
\newif\if@restonecol
\makeatother

\usepackage[boxed, ruled, vlined, noend, linesnumbered]{algorithm2e}
\SetKwRepeat{Do}{do}{while}

\newenvironment{changemargin}[2]{\begin{list}{}{
	\setlength{\topsep}{0pt}\setlength{\leftmargin}{0pt}
	\setlength{\rightmargin}{0pt}
	\setlength{\listparindent}{\parindent}
	\setlength{\itemindent}{\parindent}
	\setlength{\parsep}{0pt plus 1pt}
	\addtolength{\leftmargin}{#1}\addtolength{\rightmargin}{#2}
	}\item}
	{\end{list}}

\usepackage[first=0,last=9]{lcg}
\usepackage{colortbl}
\definecolor{Gray}{gray}{0.8}
\colorlet{Red}{red!10!white}
\colorlet{Blue}{blue!10!white}

\usepackage{hyperref}

\newtcolorbox{mtbox}[1]{left=0.25mm, right=0.25mm, top=0.25mm, bottom=0.25mm, sharp corners, colframe=blue!50!black, boxrule=0.5pt, title={#1}, fonttitle=\bfseries, coltitle=blue!50!black, attach title to upper={\ --\ }}

\newcommand{\autoremark}[1]{
  \vspace{-3pt}
  \stepcounter{remarkcounter}
  \begin{mtbox}{\small Remark\,\arabic{remarkcounter}}
    {\small #1} 
  \end{mtbox}
  \vspace{-3pt}
}

\usepackage{scalerel}[2016/12/29]

\makeatletter
\providecommand{\leadsfrom}{%
  \mathrel{\mathpalette\reflect@squig\relax}%
}
\newcommand{\reflect@squig}[2]{%
  \reflectbox{$\m@th#1\leadsto$}%
}
\makeatother

\def\eqref#1{equation~\ref{#1}}

\def\1{\bm{1}}

\DeclareMathAlphabet{\mathsfit}{\encodingdefault}{\sfdefault}{m}{sl}
\SetMathAlphabet{\mathsfit}{bold}{\encodingdefault}{\sfdefault}{bx}{n}

\newcommand{\lib}{{\sc CopyrightMeter}\xspace}

\newcommand{\minitab}[2][l]{\begin{tabular}{#1}#2\end{tabular}}
\usepackage{enumitem}
\setlist[itemize]{leftmargin=*}

\newcommand{\tbd}[1]{\textcolor{black}{#1}}
\newcommand{\naen}[1]{\textcolor{black}{#1}}
\newcommand{\minxi}[1]{\textcolor{black}{#1}}
\newcommand{\wenjie}[1]{\textcolor{black}{#1}}
\newcommand{\jiang}[1]{\textcolor{black}{#1}}

\definecolor{darkgreen}{RGB}{0, 155, 55}
\newcommand{\jc}[1]{\textcolor{black}{#1}}

\newif\ifshowcontent
\showcontentfalse
\newcommand{\hidecontent}[1]{\ifshowcontent #1 \fi}

\usepackage{amsthm}

\newcounter{remarkcounter}
\newcolumntype{M}[1]{>{\centering\arraybackslash}p{#1}}

\ifCLASSOPTIONcompsoc
  \usepackage[nocompress]{cite}
\else
  \usepackage{cite}
\fi
\ifCLASSINFOpdf
\else
\fi
\begin{document}
%
\title{\textsc{\textbf{CopyrightMeter}}: Revisiting Copyright Protection in Text-to-image Models} 


\author{
\IEEEauthorblockN{
Naen Xu\IEEEauthorrefmark{2}$^{*}$, 
Changjiang Li\IEEEauthorrefmark{4}$^{*}$\thanks{Naen Xu and Changjiang Li are the co-first authors. Tianyu Du is the corresponding author.}, 
Tianyu Du\IEEEauthorrefmark{2}(\ding{41}), 
Minxi Li\IEEEauthorrefmark{2}, 
Wenjie Luo\IEEEauthorrefmark{2}, 
Jiacheng Liang\IEEEauthorrefmark{4},
Yuyuan Li\IEEEauthorrefmark{5}, \\
Xuhong Zhang\IEEEauthorrefmark{2}, 
Meng Han\IEEEauthorrefmark{2}, 
Jianwei Yin\IEEEauthorrefmark{2}, 
Ting Wang\IEEEauthorrefmark{4}
}

\IEEEauthorblockA{
  \IEEEauthorrefmark{2}Zhejiang University,
  \IEEEauthorrefmark{4}Stony Brook University, 
  \IEEEauthorrefmark{5}Hangzhou Dianzi University \\
}

\IEEEauthorblockA{
  E-mails: \{xunaen, zjradty, breathing, zhangxuhong, mhan, zjuyjw\}@zju.edu.cn, \\
  meet.cjli@gmail.com, liminxi694@gmail.com, ljcpro@outlook.com, y2li@hdu.edu.cn, inbox.ting@gmail.com
}
}


%


\maketitle


\begin{abstract}
\jiang{Text-to-image diffusion models have emerged as powerful tools for generating high-quality images from textual descriptions. However, their increasing popularity has raised significant copyright concerns, as these models can be misused to reproduce copyrighted content without authorization. In response, recent studies have proposed various copyright protection methods, including adversarial perturbation, concept erasure, and watermarking techniques. However, their effectiveness and robustness against advanced attacks remain largely unexplored. Moreover, the lack of unified evaluation frameworks has hindered systematic comparison and fair assessment of different approaches.}

\jiang{To bridge this gap, we systematize existing copyright protection methods and attacks, providing a unified taxonomy of their design spaces. We then develop \textsc{\textbf{CopyrightMeter}}, a unified evaluation framework that incorporates 17 state-of-the-art protections and 16 representative attacks. Leveraging \textsc{\textbf{CopyrightMeter}}, we comprehensively evaluate protection methods across multiple dimensions, thereby uncovering how different design choices impact fidelity, efficacy, and resilience under attacks. Our analysis reveals several key findings: 
(i) most protections (16/17) are not resilient against attacks;
(ii) the ``best'' protection varies depending on the target priority;
(iii) more advanced attacks significantly promote the upgrading of protections.
These insights provide concrete guidance for developing more robust protection methods, while its unified evaluation protocol establishes a standard benchmark for future copyright protection research in text-to-image generation.}

\end{abstract}


%
\IEEEpeerreviewmaketitle

\section{Introduction}
Recent advances in text-to-image diffusion models (T2I DMs), such as Stable Diffusion (SD) \cite{fernandez2023stable}, DALL·E 3 \cite{dalle3}, and Imagen \cite{saharia2022photorealistic}, have revolutionized digital content creation by generating high-quality images from textual descriptions.
While these models foster creativity by producing art and realistic scenes, they also raise significant copyright concerns\cite{property}.  
\naen{Fine-tuning pre-trained models on specialized datasets allows them to mimic specific themes such as distinct art styles, which can lead to unauthorized reproductions\cite{ruiz2023dreambooth,kumari2023multi}. Artists are increasingly worried that their unique styles could be copied without permission, resulting in potential copyright infringement\cite{shan2023glaze}. Furthermore, models trained on extensive datasets may produce images that closely resemble the style or content of specific artists, even if the artist or their creations are not directly referenced in the prompt\cite{samuelson2023generative}.} 
As these AI-driven technologies evolve, it is crucial to balance innovation with the protection of creators' rights, as many artists fear that their unique art styles could be easily copied, potentially drawing customers away\cite{melissa2023this}. 

\naen{The urgent need to safeguard digital intellectual property leads to the development of three main protection categories: 
(\textit{i}) \textit{Obfuscation Processing}, which preprocesses data before release online to prevent unauthorized use, often using adversarial perturbation to confuse AI models while preserving content for normal users\cite{salman2023raising,shan2023glaze}.
(\textit{ii}) \textit{Model Sanitization}, which modifies pre-trained DMs to remove or alter protected copyright elements before public deployment\cite{gandikota2023erasing,zhang2024forget}.
(\textit{iii}) \textit{Digital Watermarking}, which embeds invisible identifiers in AI-generated content to assert copyright ownership and support effective content management\cite{cuidiffusionshield,fernandez2023stable,wen2023tree}.
}

\naen{Given the significance of these protection mechanisms, recent studies have raised concerns about their effectiveness and robustness\cite{cao2024impress,pham2023circumventing,li2023towards,jiang2023evading}. }
\tbd{This has led to several critical questions:
\textbf{RQ\textsubscript{1}} -- \textit{What are the strengths and limitations of different protection mechanisms, especially their robustness against attacks?} 
\textbf{RQ\textsubscript{2}} -- \textit{What are the best practices for copyright protection even in adversarial and envolving environments?} 
\textbf{RQ\textsubscript{3}} -- \textit{How can existing copyright protection methods be further improved?} 
}

{\hidecontent{\jiang{attacks?}}}

Despite their importance for understanding and improving copyright protection, these questions are under-explored due to the following challenge. 

\begin{table*}[t]
{\footnotesize
    \centering
    \renewcommand{\arraystretch}{1} 
    \caption{Comparison of conclusions in prior and our work ($\Circle$ -- inconsistent; $\LEFTcircle$ -- partially inconsistent; $\CIRCLE$ -- consistent).}
    \setlength{\tabcolsep}{2pt}
    \resizebox{\linewidth}{!}{
    \begin{tabular}{m{6.5cm}|m{5.8cm}|m{6.9cm}|m{1.4cm}}
    
    \multicolumn{1}{c|}{\bf Previous conclusion} & \multicolumn{1}{c|}{\bf Refined conclusion \naen{in this paper}} & \multicolumn{1}{c|}{\bf Explanation} & \multicolumn{1}{c}{\bf Consistency} \\
    \hline
    \hline
    In Obfuscation Processing, Mist shows strong effectiveness against various noise purification methods, including under the SOTA online platform NovelAI I2I scenario. & Mist has limited protective effectiveness against local DiffPure attacks and the latest version of NovelAI — NAI Diffusion Anime. &The original protection may lose resilience as new attacks circumvent current protections, rendering previous methods vulnerable.&\makecell[c]{$\LEFTcircle$ \\ (Sec \ref{sec:op} \\ and \ref{sec:realworld})} \\
    \hline
    In Model Sanitization, FMN\cite{zhang2024forget}, ESD\cite{gandikota2023erasing}, UCE\cite{gandikota2024unified}, and SLD\cite{schramowski2023safe} remove a copyright concept while preserving the model's ability to generate images without it. & All Model Sanitization methods maintain unrelated concepts without copyright concepts well.&Despite removing explicit copyright concepts, these methods ensure that the model retains its ability to generate irrelevant images, preserving its utility and effectiveness. & \makecell[c]{$\CIRCLE$ \\ (Sec \ref{sec:ms})}\\
    \hline
    In Model Sanitization, ESD permanently removes concepts from DMs, rather than modifying outputs in inference, so it cannot be circumvented even if model weights are accessible.& Model Sanitization methods are vulnerable to concept recovery methods such as DreamBooth, Text Inversion, Concept Inversion, or even model-weights-free approaches like Ring-A-Bell. &The training dataset of DM, such as LAION, contains images with varying content, and it is almost impossible to remove elements with copyright concepts permanently. &\makecell[c]{$\Circle$ \\ (Sec \ref{sec:ms})} \\
    \hline
    In Digital Watermarking, the techniques Diag, StabSig, and GShare demonstrate relative resilience against Watermark Removal attacks.& Regarding attack resilience, Diag exhibits vulnerability to Blur attacks, StabSig is vulnerable to Rotate, Blur, VAE, and DiffPure attacks, and GShade demonstrates vulnerability to Rotate attacks.& The vulnerability of Diag to Blur attack is attributed to different datasets, as the original paper employs the Pokemon dataset. Besides, StabSig and GShade are vulnerable to specific attacks not covered in the original paper.& \makecell[c]{$\LEFTcircle$ \\ (Sec \ref{sec:dw})}\\

    
    \end{tabular}
    }
    \label{tab:cmp}
}
\end{table*}




\ul{Non-holistic evaluations} -- Existing studies often lack comprehensive evaluation 
of protections and attacks  \cite{vsarvcevic2024u, ren2024copyright}, focus narrowly on limited perspectives, such as \cite{pham2023circumventing} focuses solely on model sanitization against textual inversion, without providing a holistic evaluation. Moreover, many rely on limited metrics, failing to fully capture the characteristics and impacts of the protections being evaluated.

\ul{Non-unified framework} -- Inconsistent datasets and DM versions across studies lead to evaluations under varying conditions, making comparision challenging. For example, Glaze \cite{shan2023glaze} and Mist \cite{liang2023mist} are evaluated with different SD versions, complicating direct comparisons of evaluations.

\ul{Outdated evaluations} -- 
While new attacks quickly lead to updated protections to bolster security, many studies focus solely on older protection methods, missing recent developments.
For instance, \cite{honig2024adversarial} evaluated only the original Mist system as reported by \cite{liang2023mist}, without considering the updated Mist v2 system described by\cite{zheng2023understanding}.

To solve existing issues, we introduce a systematic taxonomy for copyright protection methods and develop \lib, a systematic framework for evaluating them across different dimensions, including fidelity, efficacy, and resilience\tbd{: fidelity evaluate how protected content retains its original quality; efficacy measures the protection method's effectiveness in preventing unauthorized use or mimicry; and resilience indicates the method's ability to withstand attacks.}  By reviewing literature and evaluating current practices, our study provides insights into challenges and opportunities, guiding policymakers, content creators, and technologists striving to navigate the complex interplay between copyright law and technological advancement. Our contributions are summarized in three major aspects:

\textbf{Framework -- } \jiang{We develop \lib, the first unified framework for extensively evaluating copyright protection in T2I DMs. It integrates 17 protection methods,  16 representative attacks, and 10 key metrics for in-depth analysis of these methods. We plan to open source \lib to facilitate copyright protection research and encourage the community to contribute more techniques.}


\textbf{Evaluation -- } Leveraging \lib, we explore the landscape of copyright protection in T2I DMs, conducting a systematic study of existing protections and attacks, uncovering key insights that challenge prior conclusions, as summarized in Table \ref{tab:cmp}. Our findings reveal that different protections manifest delicate trade-offs among fidelity, efficacy, and resilience.
For instance, Mist achieves strong protection against mimicry but slightly compromises fidelity; ESD shows high efficacy but relatively weak resilience; ZoDiac and GShade have high fidelity and efficacy, but are less resilient to attacks.  
These observations indicate the importance of using comprehensive metrics to evaluate copyright protections, and suggest the optimal practices of applying them under different settings.


\textbf{Exploration -- } We further explore improving existing protections, leading to several critical insights including 
(\textit{i}) the generalizability of various copyright protection methods differs significantly; 
(\textit{ii}) in scenarios prioritizing efficiency, inference-guiding \textsc{Ms} are preferred to model fine-tuning \textsc{Ms};
(\textit{iii}) the ongoing arms race between protections and attacks promotes the development of more advanced protections.
We envision that the \lib platform and our findings will facilitate future research on copyright protection and shed light on designing and building T2I DMs in a more trustworthy manner.

\section{Background}

\begin{figure*}[t]
    \centering
    \includegraphics[width=177mm]{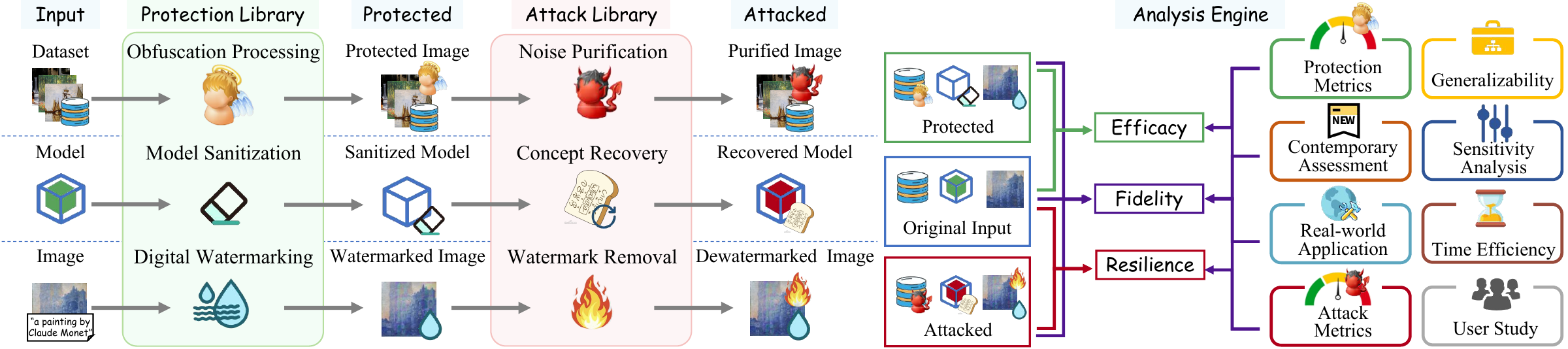}
    \caption{Overall system design of \lib.}
    \label{fig:framework}
\end{figure*}

\subsection{Text-to-image Diffusion Models}

Diffusion models are a class of generative models that transform random noise into coherent data through a forward step that gradually adds noise to data and a reverse step that denoises it to recover the original data distribution. Our study focuses on the latent diffusion models (LDMs) for their strong performance and low computational costs. 

Text-to-image diffusion models (T2I DMs) generate images from textual descriptions 
by learning to reverse the noise addition process guided by text. 
A notable open-source T2I DM example is Stable Diffusion (SD). Given a text prompt, it generates an image that reflects the specified semantic features, involves two key components:

{\bf Conditioning on Textual Descriptions} --
The reverse diffusion process is guided by textual descriptions, which are embedded into a high-dimensional vector using transformer-based models or other deep learning architectures. This vector informs each step of the reverse diffusion to align the generated image with the text.

{\bf Training Objective} -- 
T2I DMs are trained to predict and remove noise at each step, guiding image generation to match text prompts. This is achieved by minimizing the difference between actual and predicted noise:
\begin{equation}
\mathcal{L}_{DM}(\theta) = \mathbb{E}_{x_0, \epsilon, t, y} \left[ \| \epsilon - \epsilon_{\theta}(x_t, y, t) \|^2 \right]
\end{equation}
where $\epsilon$ is the noise vector, and $\epsilon_{\theta}(x_t, y, t)$ is the model's estimate of the noise, conditioned on the noisy image $x_t$, the textual description $y$, and the timestep $t$.

Beyond these design aspects, T2I DMs have driven advancements in generative AI across content creation, design, education, and entertainment, bridging the gap between textual descriptions and visual content. \naen{T2I can also be fine-tuned with tools like DreamBooth, which enables them to mimic specific visual styles or objects by training on a few reference images, thus allowing the model to produce images closely resembling these reference examples.}

\subsection{\tbd{Copyright Protection in Text-to-Image Models}}

\begin{figure}[t]
    \centering
    \includegraphics[width=85mm]{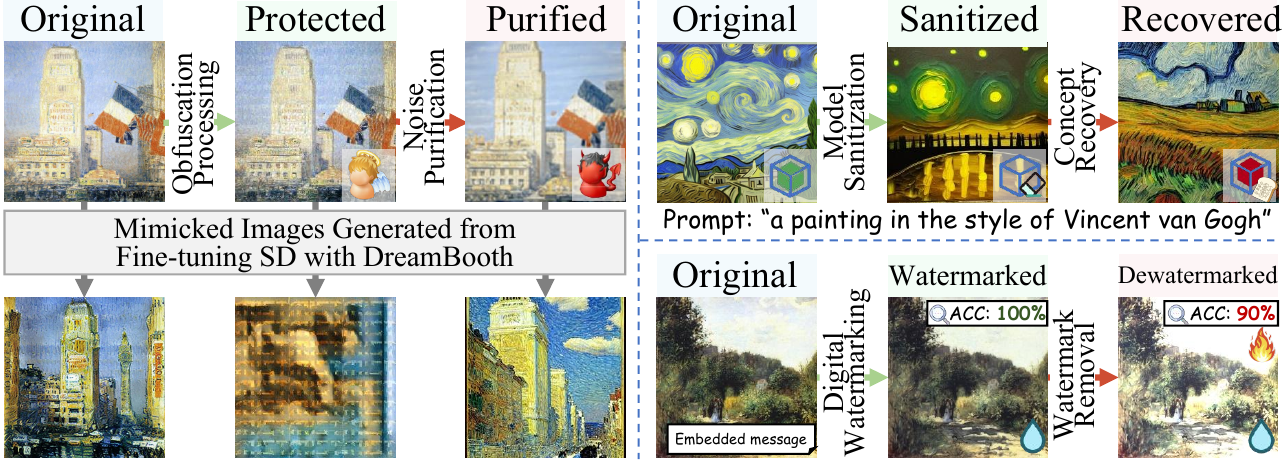}
    \caption{Examples of existing copyright protections and attacks integrated in \lib.}
    \label{fig:example}
\end{figure}



In T2I DMs, copyright protection is a critical concern. The central challenge is ensuring that images generated from the model do not resemble copyrighted images. \naen{Techniques like DreamBooth fine-tuning on DM allow models to mimic specific copyrighted content, while DMs may also inadvertently produce similar works. 
Another significant challenge is ensuring }generated images can be traced back to their copyrighted sources. Conversely, the associated attacks aim to exploit these models for unauthorized purposes. The two primary attack methods involve generating an image that matches a specific, potentially copyrighted image or manipulating the generated image to make it untraceable. We will formalize and discuss the prominent categories of protection and corresponding attack methods in Section \ref{sec:tanonomy}.

\section{Taxonomy}
\label{sec:tanonomy}

\naen{In this section, we provide a holistic overview of various copyright protection and attack methods.} As depicted in Figure \ref{fig:framework}, we divide copyright protection into three categories: \textit{Obfuscation Processing} (\textsc{Op}), \textit{Model Sanitization} (\textsc{Ms}), and \textit{Digital Watermarking} (\textsc{Dw}). Correspondingly, we identify three attack categories: \textit{Noise Purification} (\textsc{Np}), \textit{Concept Recovery} (\textsc{Cr}), and \textit{Watermark Removal} (\textsc{Wr}).
\naen{Table \ref{tab:overview-method} presents the definitions and detailed methods of these protections and their corresponding attacks. 
Figure \ref{fig:framework} shows the overall system design of \lib, while Figure \ref{fig:example} provides specific examples of copyright protection and attack scenarios.} 
We will briefly introduce each category in the subsequent sections. For convenience, we summarize the acronyms and notations in Table \ref{tab:symbols-and-notations}.

\begin{table}[t]
{\footnotesize
    \centering
    \renewcommand{\arraystretch}{1}
    \caption{Overview of copyright protections and attacks.} 
    \setlength{\tabcolsep}{2pt}
    
    \resizebox{\linewidth}{!}{
    \begin{tabular}{c|p{9.8cm}}
    
    \textbf{Category} & \multicolumn{1}{c}{\textbf{Definition and Methods}} \\
    \hline
    \hline
    \cellcolor{gray!20}Obfuscation 
        & \cellcolor{gray!20}\textbf{Definition:} Add adversarial perturbations on images to avoid image mimicry. \\
        \cellcolor{gray!20}Processing & \cellcolor{gray!20}\textbf{Methods:} AdvDM \cite{liang2023adversarial}, Mist \cite{liang2023mist}, Glaze \cite{shan2023glaze}, PGuard \cite{salman2023raising}, AntiDB \cite{van2023anti}. \\
    \hline
    \multirow{2}{*}{\minitab[c]{Noise\\Purification}} 
        & \textbf{Definition:} Purify the protected images to nullify adversarial perturbations. \\
        & \textbf{Methods:} JPEG \cite{wallace1991jpeg}, Quant \cite{heckbert1982color}, TVM \cite{chambolle2004algorithm}, IMPRESS \cite{cao2024impress}, DiffPure \cite{nie2022diffusion}. \\
    \hline
    \cellcolor{gray!20}Model 
        & \cellcolor{gray!20}\textbf{Definition:} Prevent DM from generating images containing specific concept. \\
        \cellcolor{gray!20}Sanitization& \cellcolor{gray!20}\textbf{Methods:} FMN \cite{zhang2024forget}, ESD \cite{gandikota2023erasing}, AC \cite{kumari2023ablating}, UCE \cite{gandikota2024unified}, NP \cite{stable_diffusion_webui_negative_prompt}, SLD \cite{schramowski2023safe}. \\
    \hline
    \multirow{2}{*}{\minitab[c]{Concept\\Recovery}}
        & \textbf{Definition:} Retrieve the eliminated concept to recover the content generation. \\
        & \textbf{Methods:} LoRA \cite{hu2021lora}, DB \cite{ruiz2023dreambooth}, TI \cite{gal2022image}, CI \cite{pham2023circumventing}, RB \cite{tsai2023ring}. \\
    \hline
    \cellcolor{gray!20}Digital
        & \cellcolor{gray!20}\textbf{Definition:} Embed DM-based watermark into image generation. \\
        \cellcolor{gray!20}Watermarking& \cellcolor{gray!20}\textbf{Methods:} DShield \cite{cuidiffusionshield}, Diag \cite{wang2023diagnosis}, StabSig \cite{fernandez2023stable}, ZoDiac \cite{zhang2024robust}, TR \cite{wen2023tree}, GShade \cite{yang2024gaussian}. \\
    \hline
    \multirow{2}{*}{\minitab[c]{Watermark\\Removal}} 
        & \textbf{Definition:} Tamper with images to remove watermark. \\
        & \textbf{Methods:} Bright \cite{yang2024gaussian}, Rotate \cite{yang2024gaussian}, Crop \cite{yang2024gaussian}, Blur \cite{hosam2019attacking}, VAE \cite{cheng2020learned}, DiffPure \cite{nie2022diffusion}. \\
    \end{tabular}
    }
    \label{tab:overview-method}
}
\end{table}

\begin{table}[tp]
{\footnotesize
    \centering
    \renewcommand{\arraystretch}{0.8}
    \caption{Acronyms and notations.}
    \setlength{\tabcolsep}{5pt}
    \resizebox{\linewidth}{!}{
    \begin{tabular}{>{\centering\arraybackslash}p{0.2cm}|>{\centering\arraybackslash}p{1.4cm}|>{\raggedright\arraybackslash}p{7.2cm}}
    \multicolumn{2}{c|}{\textbf{Notation}} & \multicolumn{1}{c}{\textbf{Definition}} \\ 
    \hline
    \hline  
    \multirow{5}{*}{\rotatebox{90}{General}} & $x$ & original copyrighted image to be protected \\
    & $\theta$ & Diffusion Model (DM)'s weights \\
    & $D_{p}(\cdot,\cdot)$ & pixel distance between images \\ 
    & $D_{z}(\cdot,\cdot)$ & latent space distance between images\\ 
    & $\epsilon$ & upper bound of pixel distance between two images \\
    
    \hdashline
    \multirow{5}{*}{\rotatebox{90}{\textsc{Op} \& \textsc{Np}}} & $\delta$ & perturbation introduced during obfuscation processing (\textsc{Op}) \\ 
    & $x_{\text{t}}$ & the chosen dissimilar target image differ from $x$ in \textsc{Op}  \\  
    & $x_\text{pro}$ & protected image with perturbation applied in \textsc{Op}  \\
    & $\tau$ & transformation applied in noise purification (\textsc{Np}) to remove $\delta$\\
    & $x_{\text{pur}}$ & purified image after $\tau$ in \textsc{Np} \\

    \hdashline
    \multirow{6}{*}{\rotatebox{90}{\textsc{Ms} \& \textsc{Cr}}} & $C$ & set of all possible concepts \\
    & $c_{\text{cr}}$ & copyright concept for model sanitization (\textsc{Ms}) \\
    & $c_{\text{$\varnothing$}}$ & a specific unrelated concept that excludes $c_{\text{cr}}$\\ 
    & $c_{\text{ref}}$ & concept in reference image similar to copyrighted image $x$ \\
    & $p(x|c)$ & image generation distribution by DM given concept $c$ \\
    & \tbd{$D_{KL}(\cdot \parallel \cdot)$} &  \tbd{the divergence between the two image output distributions} \\

    \hdashline
    \multirow{8}{*}{\rotatebox{90}{\textsc{Dw} \& \textsc{Wr}}} & $m$ & original watermarked message embedded into $x$ \\
    & $w$ & watermark embedding function \\ 
    & $e$ & watermark extraction function \\ 
    & $x_{\text{wm}}$ & watermarked image with message $m$ via $w$ \\
    & $m_{\text{wm}}$ & extracted message from $x_\text{wm}$ \\
    & $x_{\text{wr}}$ & image after watermark removal \\
    & $D_{t}(\cdot,\cdot)$ & text distance between two watermarked messages \\ 
    \end{tabular}
    }
    \label{tab:symbols-and-notations}
}
\end{table}




\subsection{Protection Schemes}
This subsection contains survey-style descriptions of the investigated copyright protection schemes. Table \ref{tab:protection-summary} shows the copyright protection methods and detailed characteristics.



\begin{table*}[t]
\centering
{\scriptsize 
    
    \renewcommand{\arraystretch}{0.8}
    \caption{Summary of copyright protection methods in \lib.} 
    \setlength{\tabcolsep}{5pt}
    \centering
    \begin{adjustbox}{width=\textwidth}
    \begin{tabular}{r|c|c|p{.8cm}<{\centering}|p{.8cm}<{\centering}|p{.8cm}<{\centering}|p{.8cm}<{\centering}|c|c}
    {\bf \multirow{2}{*}{\minitab[c]{Protection}}} & {\bf \multirow{2}{*}{\minitab[c]{Category}}} & {\bf \multirow{2}{*}{\minitab[c]{Auxiliary Guidance}}} & \multicolumn{2}{c|}{\bf Distortion} & \multicolumn{2}{c|}{\bf Scenario} & {\bf \multirow{2}{*}{\minitab[c]{Main Application}}} & {\bf \multirow{2}{*}{\minitab[c]{Implementer}}} \\ 
    \cline{4-7} 
    & & & {\bf Sem.} & {\bf Graph.} & {\bf I2I} & {\bf T2I} & & \\ 
    \hline
    \hline
    
    AdvDM \cite{liang2023adversarial}& \multirow{5}{*}{\minitab[c]{Obfuscation\\Processing}} & Diffusion Model  & $\checkmark$ &  $\times$ & $\checkmark$ &  $\checkmark $ & Unauthorized style mimicry  & \multirow{5}{*}{\minitab[c]{Data Owner}}\\
    Mist \cite{liang2023mist}& & Image Encoder \& Diffusion &  $\checkmark$ &  $\checkmark $ & $\checkmark$ &  $\checkmark$ & Unauthorized style mimicry & \\
    Glaze \cite{shan2023glaze}&  & Image Encoder & $\checkmark$ &  $\times$ & $\checkmark$ &  $\checkmark$ & Unauthorized style mimicry &  \\ 
    PGuard \cite{salman2023raising}& & Image Encoder & $\times$ & $\checkmark$ & $\checkmark$ &  $\times$ & Unauthorized editing & \\
    AntiDB \cite{van2023anti}& & Diffusion Model & $\checkmark$ &  $\checkmark$ & $\times$ &  $\checkmark$ & Unauthorized image mimicry& \\
    
    \hline
    FMN \cite{zhang2024forget}& \multirow{6}{*}{\minitab[c]{Model\\Sanitization}} & Model Fine-tuning & $\checkmark$ & $\times$ & $\times$ &  $\checkmark$ & Identity, object, or style & \multirow{6}{*}{\minitab[c]{Model Provider}} \\
    ESD \cite{gandikota2023erasing}& & Model Fine-tuning &$\checkmark$ & $\times$ & $\times$ &  $\checkmark$ & Style, explicit content, or object & \\  
    AC \cite{kumari2023ablating}& & Textual Inversion &$\checkmark$ & $\times$ & $\times$ &  $\checkmark$ & Instance, style, or memorized images & \\
    UCE \cite{gandikota2024unified}& & Textual Inversion &$\checkmark$ & $\times$ & $\times$ &  $\checkmark$ & Artist or objects   & \\ 
    NP \cite{stable_diffusion_webui_negative_prompt}& & Inference Guiding &  $\checkmark$ & $\times$ & $\times$ &  $\checkmark$ & Specific concepts or features  & \\
    SLD \cite{schramowski2023safe}& &Inference Guiding & $\checkmark$ & $\times$ & $\times$ &  $\checkmark$ & Inappropriate concept & \\
    \hline
    
    DShield \cite{cuidiffusionshield}& \multirow{6}{*}{\minitab[c]{Digital\\Watermarking}} &Model Fine-tuning & $\times$ & $\checkmark$ & $\checkmark$ &  $\times$ &  Multi-bit watermark for existing images& \multirow{6}{*}{\minitab[c]{Works Publisher}} \\ 
    Diag \cite{wang2023diagnosis}& & Model Fine-tuning & $\times$ & $\checkmark$ & $\checkmark$ &  $\times$ & Zero-bit watermark for existing images& \\   
    StabSig \cite{fernandez2023stable}& & Model Fine-tuning &$\times$  & $\checkmark$ & $\times$ &  $\checkmark$ & Multi-bit watermark for generating images& \\
    TR \cite{wen2023tree}& & Latent Space Modifying& $\times$ &$\checkmark$ & $\times$ &  $\checkmark$ & Zero-bit watermark for generating images& \\
    ZoDiac \cite{zhang2024robust}& & Latent Space Modifying& $\times$ & $\checkmark$ & $\checkmark$ &  $\times$ & Zero-bit watermark for existing images& \\  
    GShade \cite{yang2024gaussian}& & Latent Space Modifying & $\times$ & $\checkmark$ & $\times$ &  $\checkmark$ & Multi-bit watermark for generating images& \\
    
    \end{tabular}
    \end{adjustbox}
    \label{tab:protection-summary}
    \begin{flushleft}
    {\footnotesize
    \textit{Note:} \tbd{Auxiliary Guidance -- model components integrated for perturbation optimization in \textsc{Op}, or methods used in \textsc{Ms} and \textsc{Dw}.} Sem -- the semantic-distortion-based method, Graph -- the graphical-distortion-based method, I2I -- image-to-image generation; T2I -- text-to-image generation.
    }
    \end{flushleft}

}
\end{table*}


\subsubsection{Obfuscation Processing (\textsc{Op})} 

This approach introduces protective perturbations into copyrighted images to prevent replication from T2I DMs. When these protected images are used as training or reference data (\textit{e.g.}, in image-to-image transformation), they mislead DMs that aim to replicate the originals, thereby protecting data owners from unauthorized replication and misuse of their data.

\textbf{Formalization} -- \jiang{Given a copyrighted image $x$, the aim is to create a protected image $x_{\text{pro}}$ by adding a carefully crafted perturbation $\delta$, such that $x_{\text{pro}} = x + \delta$. This perturbation $\delta$ is designed to either maximize the latent space distance between $x_{\text{pro}}$ and $x$ (untargeted protection) or minimize the latent space similarity between $x_{\text{pro}}$ and a deliberately chosen dissimilar target image $x_t$ (targeted protection).
Additionally, to ensure the perturbation remains inconspicuous, the pixel distance between $x$ and $x_{\text{pro}}$ should be }\naen{constrained by an upper bound $\epsilon$}\jiang{, maintaining the visual fidelity of the protected image. This can be formatted as:  } 
\small
\begin{equation}
\max_{\delta} D_{z}(x, x_{\text{pro}}) \; \text{or} \; \min_{\delta} D_{z}(x_{\text{pro}}, x_t), \text{s.t. } D_{p}(x, x_{\text{pro}}) \leq \epsilon.
\end{equation}
\normalsize

\textbf{Approaches} -- \jiang{Since all methods maintain visual similarity by ensuring the perturbation $\delta$ }  \jiang{maintain a small} \jiang{pixel space distance between $x$ and $x_{\text{pro}}$, we omit this commonality and focus solely on the unique protection concepts of each method.}  
AdvDM \cite{liang2023adversarial} optimizes $\delta$ to maximize the diffusion training loss and increase the latent noise vector's distance of $x_\text{pro}$ and $x$. 
Based on AdvDM, Mist \cite{liang2023mist} optimizes $\delta$ to maximize distance both in the latent noise vector and latent encoded representation. 
\jiang{Glaze \cite{shan2023glaze} optimizes $\delta$ by adjusting it to approach $x_t$ with a specific style, aiming to minimize $D_{z}(x_\text{pro}, x_t)$. 
PhotoGuard (PGuard) \cite{salman2023raising} using two schemes -- using either the encoder or the entire diffusion process to optimize $\delta$ to minimize $D_{z}(x_\text{pro}, x_t)$ in the latent space of encoder and LDM, respectively.}
\naen{Anti-DreamBooth (AntiDB) \cite{van2023anti}} optimizes \( \delta \) to minimize DM's generation ability by making $x$ difficult to reconstruct from $x_\text{pro}$. 

\subsubsection{Model Sanitization (\textsc{Ms})} This approach is designed for model providers by guiding pre-trained DMs to remove copyright concepts before public deployment, ensuring that the models do not reproduce copyrighted content illegally.




\textbf{Formalization} -- Given a concept protected by copyright $c_{\text{cr}} \in C$ (where $C$ is the set of all concepts) and a specific unrelated concept $c_{\text{$\varnothing$}} \in C \setminus c_{\text{cr}}$. It shifts model's generation distribution conditioned on $c_{\text{cr}}$, denoted as $p_{\phi}(x|c_{\text{cr}})$, toward the distribution conditioned on the unrelated concept $c_{\text{$\varnothing$}}$, denoted as $p_{\phi}(x|c_{\text{$\varnothing$}})$. To measure the alignment, 
we minimize the KL divergence $D_{KL}$ between these distributions through the transformation $\phi$, the model's output distribution is adjusted to reduce its ability to generate images corresponding to $c_\text{cr}$. The objective can be formalized as:
\begin{equation}
\arg\min_{\phi}D_{KL}(p(x|c_{\text{$\varnothing$}})\parallel p_{\phi}(x|c_{\text{cr}})).
\end{equation}

\textbf{Approaches} --  Based on the difference in distribution alignment, the approaches can be categorized into two types: \textit{fine-tuning} and \textit{inference guiding} methods.

\textit{Fine-tuning} methods adjust $p_{\phi}(x|c_{\text{cr}})$ by modifying the DM's U-Net weights, targeting different components depending on the method \cite{ronneberger2015u}. 
For instance, Forget-Me-Not (FMN) \cite{zhang2024forget} fine-tunes U-Net cross-attention layers' weights to minimize the Frobenius norm of attention maps between input feature and embedding of $c_\text{cr}$, aligning $p_{\phi}(x|c_{\text{cr}})$ more closely with $p(x|c_{\text{$\varnothing$}})$.
Erased Stable Diffusion (ESD) \cite{gandikota2023erasing} fine-tunes to both cross-attention and unconditional layers to diminish $c_\text{cr}$'s influence in denoising prediction. 
Ablating Concepts (AC) \cite{kumari2023ablating} further fine-tunes U-Net weights, including projection matrices in cross-attention layers, and text transformer embedding to minimize KL divergence for a tighter alignment. 
Unified Concept Editing (UCE) \cite{gandikota2024unified} strategically modifies U-Net's cross-attention keys and values associated with text embeddings of $c_\text{cr}$ to align $p_{\phi}(x|c_{\text{cr}})$ with $p(x|c_{\text{$\varnothing$}})$ while preserving unrelated concepts $c_{\text{$\varnothing$}}$. 

\textit{Inference guiding} methods adjust the sampling process without altering model weights. In SD, each sampling step involves conditional and unconditional denoising. The final noise prediction is derived by taking the difference between these two samplings. 
Negative Prompt (NP) \cite{stable_diffusion_webui_negative_prompt} replaces unconditional noise prediction with noise conditioned on $c_\text{cr}$, guiding diffusion away from the $c_\text{cr}$.
Safe Latent Diffusion (SLD) \cite{schramowski2023safe} adds a safety guidance term, further shifting the distribution away from $p_{\phi}(x|c_{\text{cr}})$. 

\subsubsection{Digital Watermarking (\textsc{Dw})}
This approach embeds invisible messages in images to trace image origins and verify copyright. 
Unlike traditional post-hoc watermarks \cite{shih2017digital,zhang2019robust} applied after image generation and do not involve DMs, we discussed watermarks in the generation process of DMs.
This can be achieved by embedding watermarks directly in the training data and fine-tuning the DM, or by modifying latent vectors to impact the generation of images.





\begin{table*}[t]
\centering
{\scriptsize
    
    \renewcommand{\arraystretch}{0.8}
    \centering
    \setlength{\tabcolsep}{5pt}

    \caption{Summary of copyright attack methods in \lib.}
    \label{tab:attack-summary}
    \begin{adjustbox}{width=\textwidth}
    \begin{tabular}{r|c|c|c|p{.8cm}<{\centering}|p{.8cm}<{\centering}|p{.8cm}<{\centering}|p{.8cm}<{\centering}|p{.8cm}<{\centering}|c|c} 
    {\bf \multirow{2}{*}{\minitab[c]{Attack}}} & {\bf \multirow{2}{*}{\minitab[c]{Category}}} & {\bf \multirow{2}{*}{\minitab[c]{Approach\\Type}}} & {\bf \multirow{2}{*}{\minitab[c]{Methodology}}} & \multicolumn{3}{c|}{\bf Accessibility} & \multicolumn{2}{c|}{\bf Scenario} & {\bf \multirow{2}{*}{\minitab[c]{Target}}} & {\bf \multirow{2}{*}{\minitab[c]{Capability\\of Adversary}}} \\ 
    \cline{5-9} 
    & & & & {\bf Text} & {\bf Image} & {\bf Model} & {\bf I2I} & {\bf T2I} & & \\ 
    \hline
    \hline

    JPEG \cite{wallace1991jpeg}& \multirow{5}{*}{\minitab[c]{Noise\\Purification}} & Empirical & Data Compression & $\times$ &$\checkmark$ & $\times$ & $\checkmark$ & $\times$ &Lossy Compression  & \multirow{5}{*}{\minitab[c]{Raw Data\\Availability}} \\
    Quant\cite{heckbert1982color} & & Empirical & Data Compression & $\times$ & $\checkmark$ & $\times$ &$\checkmark$ & $\times$ &Lossy Compression  &  \\
    TVM \cite{chambolle2004algorithm}&  & Optimization & Denoising and Smoothing & $\times$ & $\checkmark$ & $\times$ & $\checkmark$ & $\times$ &Perturbation Purification &  \\
    IMPRESS \cite{cao2024impress}& & Optimization & Denoising and Smoothing & $\times$ &$\checkmark$ & $\checkmark$ & $\checkmark$ & $\times$ &Perturbation Purification & \\
    DiffPure \cite{nie2022diffusion}& & Optimization & Image Regeneration & $\times$ & $\checkmark$ & $\checkmark $ &$\checkmark$ & $\times$ &Perturbation Purification & \\
    
    \hline
    LoRA \cite{hu2021lora}&\multirow{5}{*}{\minitab[c]{Concept\\Recovery}} & Optimization & Model Fine-Tuning & $\checkmark$ &$\checkmark$ & $\checkmark$ & $\checkmark$ & $\checkmark$ &Personalizing Generation & \multirow{5}{*}{\minitab[c]{Model Weights\\Availability}} \\
    DB \cite{ruiz2023dreambooth}& & Optimization & Model Fine-Tuning & $\checkmark$ & $\checkmark$ & $\checkmark$  & $\checkmark$ & $\checkmark$ &Personalizing Generation &  \\
    TI \cite{gal2022image}&  & Optimization & Model Fine-Tuning & $\checkmark$  &$\checkmark$ & $\checkmark$ &  $\checkmark$ & $\checkmark$ &Personalizing Generation&  \\
    CI \cite{pham2023circumventing}& & Optimization & Model Fine-Tuning & $\checkmark$ &$\checkmark$ & $\checkmark$ & $\checkmark$ & $\times$ & Sanitized Concepts Retrieval &  \\
    RB \cite{tsai2023ring}& & Optimization & Prompt Engineering & $\checkmark$  & $\times$ & $\times$ & $\times$ & $\checkmark$ &Sanitized Concepts Retrieval &  \\
    
    \hline
    Bright \cite{yang2024gaussian}& \multirow{6}{*}{\minitab[c]{Watermark\\Removal}} & Empirical & Image Distortion & $\times$ &$\checkmark$ & $\times$ & $\checkmark$ & $\times$ & Watermark Obscuration & \multirow{6}{*}{\minitab[c]{Final Image\\Availability}} \\
    Rotate \cite{yang2024gaussian}& & Empirical & Image Distortion & $\times$ &$\checkmark$ & $\times$ & $\checkmark$ & $\times$ &Watermark Obscuration&  \\
    Crop \cite{yang2024gaussian}& & Empirical & Image Distortion & $\times$ &$\checkmark$ & $\times$ & $\checkmark$ & $\times$ &Watermark Obscuration &  \\
    Blur \cite{hosam2019attacking}& & Empirical & Image Distortion & $\times$ &$\checkmark$ & $\times$ & $\checkmark$ & $\times$ &Watermark Obscuration &  \\
    VAE \cite{cheng2020learned}& & Optimization & Image Regeneration & $\times$ &$\checkmark$ & $\checkmark$ & $\checkmark$ & $\times$ &Image Compression &  \\
    DiffPure \cite{nie2022diffusion}& & Optimization & Image Regeneration & $\times$ &$\checkmark$ & $\checkmark$& $\checkmark$ & $\times$ &Perturbation Purification &  \\

    \end{tabular}
    \end{adjustbox}
}
\end{table*}

\textbf{Formalization} -- 
Embedding a watermark message $m$ into an image $x$ with a function $w$ results in a watermarked image $x_{\text{wm}}=w(x, m)$. An extraction function $e$ is decodes the message $m_\text{wm}=e(x_{\text{wm}})$. 
The watermarked image $x_{\text{wm}}$ should remain visually similar to the $x$, and the $m_\text{wm}$ should accurately reflect $m$. 
The goal is to find $w$ that minimizes either pixel distance  $D_p$ or latent space distance $D_z$ between $x$ and $x_{\text{wm}}$, while optionally minimizing the text discrepancy $D_\text{t}$ between $m$ and $m_\text{wm}$, depending on the specific method. This can be formalized as:
\begin{equation}
\min_{w} \left[ \alpha D_p(x, x_{\text{wm}}) + \beta D_z(x, x_{\text{wm}}) + \lambda D_t(m, m_\text{wm}) \right],
\end{equation}
where \(\alpha\), \(\beta\), and \(\lambda\) are weights that balance image quality, latent space similarity, and message accuracy, respectively. Depending on the approach, either \(D_p\) or \(D_z\) (or both) may be used, and \(D_t\) is included if relevant.

\textbf{Approaches} -- 
The following methods outline different watermark embedding processes, denoted by $w$. 
DiffusionShield (DShield) \cite{cuidiffusionshield} encodes the watermark message \(m\) as a binary sequence, embedding each bit into distinct regions of the image \(x\), with a decoder optimized to minimize the discrepancy between \(m_{\text{wm}}\) and \(m\), while controlling the \(\ell_{\infty}\)-norm to reduce the pixel distance between \(x\) and \(x_{\text{wm}}\). 
Diagnosis (Diag) \cite{wang2023diagnosis} applies a text trigger to a dataset subset, fine-tuning the model to generate $x_{\text{wm}}$, and trains a binary classifier for watermark detection.
Stable Signature (StabSig) \cite{fernandez2023stable} fine-tunes the decoder of the image generator with a binary signature, producing $x_{\text{wm}}$ while minimizing perceptual distortion $D_p(x, x_{\text{wm}})$ and message discrepancy $D_t(m, m_\text{wm})$. 
Tree-Ring (TR) \cite{wen2023tree} embeds $m$ in the Fourier space of initial noise latent vector, detectable through DDIM inversion\cite{dhariwal2021diffusion}, while minimizing $L_{1}$ distance between $m$ and $m_\text{wm}$ from the Fourier transform of the inverted noise vector.  
ZoDiac \cite{zhang2024robust} is equipped for watermarking existing images by embedding $m$ into the latent vector through DDIM inversion, incorporating Euclidean distance, SSIM loss, and Watson-VGG perceptual loss to minimize the pixel distance of $x_\text{wm}$ and $x$.
Gaussian Shading (GShade) \cite{yang2024gaussian} maps $m$ to latent
representations following a standard Gaussian distribution, aiming to preserve the distribution between $x$ and $x_\text{wm}$ for fidelity. 
\subsection{Attack Schemes}
This subsection outlines the copyright attack schemes evaluated. Table \ref{tab:attack-summary} summarizes attack methods and their detailed characteristics.




\subsubsection{Noise Purification (\textsc{Np})}
This process employs specific transformation as an attack to remove the protective perturbations added to images in \textsc{Op}, thereby evaluating the effectiveness of \textsc{Op} under attack and assessing its resilience.




\textbf{Formalization} -- Given a protected image $x_\text{pro}=x+\delta$, the adversary aims to apply a transformation $\tau$ to remove the perturbation $\delta$. These methods can be classified into two categories: 
(\textit{i}) \textit{Experience-based} methods, which use common transformations (\textit{e.g.}, JPEG compression) as $\tau$ to remove perturbation $\delta$ while having little impact on the pixels difference between $x_{\text{pro}}$ and $x_{\text{pur}}$.
(\textit{ii}) \textit{Optimization-based} methods eliminate the potential protection \(\delta\) more accurately by customizing transformations to align the latent and pixel spaces of \(x_{\text{pur}}\).
Specifically, it minimizes the pixel distance between \( x_{\text{pur}} \) and reconstructed image \( f_{\theta}(x_{\text{pur}}) \) generated from the latent representation. 
Besides, for purification fidelity, it is crucial that \(x_{\text{pur}}\) remains visually similar to the original image \(x\). 
However, 
as \(x\) is typically unavailable during attacks. Therefore, \(x_{\text{pro}}\) is used to approximate \(x\) due to the minor perturbation \(\delta\). Visual similarity is then achieved by constraining the pixel distance between \(x_{\text{pur}}\) and \(x_{\text{pro}}\).  This overall process can be formatted as:
\begin{equation}
    \min_{\tau}D_{p}(x_{\text{pur}}, f_{\theta}(x_{\text{pur}})), \text{s.t. } D_{p}(x_{\text{pro}}, x_{\text{pur}}) \leq \epsilon.
\end{equation}

\textbf{Approaches} -- 
\textit{Experience-based} methods use the following transformation as $\tau$: JPEG \cite{wallace1991jpeg} is a lossy compression algorithm that uses discrete cosine transform to remove high-frequency components from $x_\text{pro}$; 
Quantization (Quant) \cite{heckbert1982color} compresses pixel values to single discrete values.

\textit{Optimization-based} methods include: Total Variation Minimization (TVM) \cite{chambolle2004algorithm} reduces $\delta$ by minimizing unnecessary pixel intensity variations (\textit{i.e.}, gradient amplitude). 
IMPRESS \cite{cao2024impress} purifies $x_\text{pro}$ by minimizing the consistency between $x_\text{pur}$ and $f_{\theta}(x_{\text{pur}})$ while limiting the LPIPS between $x_\text{pro}$ and $x_\text{pur}$ for visual similarity; 
DiffPure \cite{nie2022diffusion} adds noise to $x_\text{pro}$ and then denoises to remove $\delta$, limiting the upper bound of pixel distance between $x_\text{pro}$ and $x_\text{pur}$.

\subsubsection{Concept Recovery (\textsc{Cr})} This process targets 
vulnerabilities to recover sanitized concepts, enabling sanitized models to generate images with copyrighted concepts, thus posing a risk of illegal replication. This evaluation assesses the resilience of sanitized models to such recovery attempts.


\textbf{Formalization} -- 
For a sanitized model with output distribution $p_{\theta}(x|c_{\text{cr}})$ aligned with unrelated concept distribution $p(x|c_{\text{$\varnothing$}})$, \textsc{Cr} aims to realign $p_{\theta}(x|c_{\text{cr}})$ to a reference distribution $p(x|c_{\text{ref}})$. This reference distribution corresponds to images containing $c_{\text{ref}}$, which are similar to the copyright content. The goal is to minimize the divergence between $p(x|c_{\text{ref}})$ and $p_{\theta}(x|c_{\text{cr}})$, enabling the sanitized model to regenerate images containing $c_\text{cr}$. This can be formatted as:
\begin{equation}
\arg\min_{\theta}D_{KL}(p(x|c_{\text{ref}})\parallel p_{\theta}(x|c_{\text{cr}})).
\end{equation}

\textbf{Approaches} -- 
These methods learn embeddings from reference images with $c_\text{ref}$ and adjust the sanitized model to realign its output distribution. LoRA \cite{hu2021lora} modifies $\theta$ using a low-rank decomposition of weight updates, efficiently fine-tuning the model to align \(p_{\theta}(x|c_{\text{cr}})\) with \(p(x|c_{\text{ref}})\). 
Similarly, DreamBooth (DB) \cite{ruiz2023dreambooth} fine-tunes models on a set of images with $c_{\text{ref}}$, embedding $c_\text{ref}$ into the model's output domain to produce images with the distribution \(p(x|c_{\text{ref}})\). 
Textual Inversion (TI) \cite{gal2022image} optimizes embedding for \(c_{\text{ref}}\) by modifying the loss function to incorporate \(c_{\text{ref}}\) during noise prediction, minimizing the discrepancy between noise predictions for generated and reference images.
Concept Inversion (CI) \cite{pham2023circumventing} learns specialized embeddings that can recover $c_\text{cr}$ for each \textsc{Ms} approach to further improve alignment with $c_\text{cr}$. 
Ring-A-Bell (RB) \cite{tsai2023ring} is a model-agnostic method that extracts holistic representations of $c_\text{cr}$ to identify prompts that might trigger unauthorized generation of copyright content. 
{\hidecontent{\jiang{TBD}\naen{$\checkmark$}}}

\subsubsection{Watermark Removal (\textsc{Wr})} To assess the resilience of \textsc{Dw} against watermark removal, this approach evaluates watermark robustness by attempting to remove them.

\textbf{Formalization} -- Given a watermarked image \(x_{\text{wm}}\), an adversary applies typical image transformation attack \(a\) to generate a watermark-removed image \(x_{\text{wr}} = a(x_{\text{wm}})\). The goal is to make the watermark undetectable while keeping \(x_{\text{wr}}\) visually similar to \(x_{\text{wm}}\). Following \cite{wen2023tree,zhao2023invisible}, 
the pixel-level distortion between \(x_{\text{wr}}\) and \(x_{\text{wm}}\) is constrained to stay below a threshold \(\epsilon\), ensuring visual similarity.
Formally, this objective is expressed as:{\hidecontent{\jiang{hard to buy}\naen{$\checkmark$}}}
\begin{equation}
D_p(x_{\text{wm}}, x_{\text{wr}}) \leq \epsilon.
\end{equation}

\textbf{Approaches} -- 
Brightness Adjustment (Bright) \cite{yang2024gaussian} adjusts the brightness of $x_\text{wm}$ to produce $x_\text{wr}$.
Image Rotation (Rotate) \cite{yang2024gaussian} rotates $x_\text{wm}$ 
to disrupt synchronization between the watermark embedder and detector. 
Random Crop (Crop) \cite{yang2024gaussian} removes portions of $x_\text{wm}$. 
Gaussian Blur (Blur) \cite{yang2024gaussian} convolves $x_\text{wm}$ with a Gaussian kernel to smooth the image and reduce watermark visibility. 
VAE-Cheng20 (VAE) \cite{cheng2020learned} compresses $x_\text{wm}$ using discretized Gaussian mixture likelihoods and attention modules to obscure the watermark. 
DiffPure \cite{nie2022diffusion} adds noise to $x_\text{wm}$, followed by DM-based denoising to remove the watermark.

\vspace{-0.15cm}
\subsection{Threat Model}
\vspace{-0.15cm}


We systematically categorize the security threats to copyright protection methods based on the adversary's objective, knowledge, and capability.

\jiang{\textbf{Adversary's objective.} In the field of text-to-image (T2I) diffusion models, adversaries aim to generate specific style/concept images. They exploit system flaws and challenge security measures to enable illegal copying and editing of images. Their objectives are multifaceted, including emulating a specific artist's style, undeterred by existing obfuscation protections, the regeneration of \naen{sanitized} concepts from purposefully sanitized models, and evading watermark detection. All these endeavors are pursued while maintaining a level of quality akin to the original copyrighted images.
 }

\jiang{\textbf{Adversary's knowledge.} Considering the variations in different protection methods, we've tailored our model of the adversary's background knowledge to capture these nuances. For obfuscation protections and digital watermarking, the adversary is capable of accessing the safeguarded or watermarked artistic images. In model sanitization, the adversary can access the sanitized model and a small set of reference images embodying the target concepts.}

\jiang{\textbf{Adversary's capability.} In a similar vein, we've adjusted our model of the adversary's capability to reflect these nuances. For obfuscation processing and digital watermarking, the adversary can modify the protected or watermarked images. In the context of model sanitization, the adversary can draw upon their knowledge of \naen{sanitized} methods to retrain \naen{sanitized} models using example images, thereby recovering the \naen{sanitized} concepts.
}

\section{Experiments}

\naen{Leveraging \lib, we conduct a systematic evaluation of existing copyright protection and attack methods, uncovering their intricate design landscape. 
}

\subsection{Experimental Setup}
\label{sec:exp-setup}
\textbf{Datasets. }
We evaluate on three datasets: WikiArt \cite{saleh2015large}, CustomConcept101 \cite{kumari2023multi} (referred to as Concept), and Person \cite{pham2023circumventing}. WikiArt contains over 42,000 artworks from 129 artists, categorized by genre (\textit{e.g.}, Impressionism). 
Concept consists of images of 101 specific concepts, each with 3 to 15 images. 
Person consists of photos of 10 distinct celebrities, with 15 images for each individual derived from the LAION dataset \cite{schuhmann2022laion}. 
For \textsc{Op}, following \cite{cao2024impress, zhang2024robust}, we use WikiArt and Concept. For \textsc{Ms}, following \cite{zhang2024forget,pham2023circumventing}, we use WikiArt and Person. For \textsc{Dw}, we use all three datasets.



\textbf{Models.} We evaluate the widely used and open-source DM implementation Stable Diffusion (SD). 
Previous studies used different versions of the SD, making it difficult to isolate the effects of copyright protection methods from model variations. We select a representative SD \cite{rombach2022high} version 1.5\footnote{\href{https://huggingface.co/runwayml/stable-diffusion-v1-5}{https://huggingface.co/runwayml/stable-diffusion-v1-5}}, the most widely downloaded version on the Hugging Face platform with a resolution of 512$\times$512 as the T2I DM in image generation experiments for a unified evaluation. 

\textbf{Metrics.} Following \cite{cao2024impress,liang2023adversarial,salman2023raising}, \lib incorporates several key metrics: 
Peak Signal-to-Noise Ratio (PSNR) quantifies the ratio between maximum possible signal power and noise; 
Structural Similarity Index Measure (SSIM) evaluates structural similarity, brightness, and contrast between two images; 
Visual Information Fidelity (VIFp) assesses image quality based on information fidelity; 
Learned Perceptual Image Patch Similarity (LPIPS) uses deep learning features for perceptual similarity assessment;
Fréchet Inception Distance (FID) measures the distribution distance between feature vectors for real and generated images;
CLIP-I and CLIP-T use CLIP model \cite{radford2021learning} to assess the image-image similarity and text-image alignment, respectively;
ACC denotes the detection accuracy of watermarks.
Except for FID and LPIPS, higher values indicate closer alignment with the reference image or corresponding text. 
Table \ref{tab:watermarking_requirements} in Appendix \ref{app:image-result} provides a detailed breakdown of these metrics across \textit{fidelity}, \textit{efficacy}, and \textit{resilience}.

\textbf{Implementation.} All experiments are conducted on a server with two Intel Xeon CPUs, 64 GB memory, a 4TB HDD, and an NVIDIA A800 GPU. Appendix \ref{app:exp-setup} details experimental setup for copyright protections and attacks.

\subsection{Obfuscation Processing Evaluation}
\label{sec:op}
\label{sec:op-evaluation}
In this subsection, we evaluate the performance of obfuscation processing (\textsc{Op}) methods to understand how different design choices affect their effectiveness. Specifically, we begin by applying the \textsc{Op} methods to generate the protected images, and then employ DreamBooth \cite{ruiz2023dreambooth} to mimic the style of these protected images. \naen{Our evaluation focuses on three aspects: \textit{fidelity}, which measures the similarity between protected and original images; \textit{efficacy}, which measures how effectively the protected images prevent mimicry; and \textit{resilience}, which examines the robustness of protection when using noise purification (\textsc{Np}) attacks to remove the perturbation.} The setup details are shown in Appendix \ref{app:protection-data}.

\begin{figure}[t]
    \centering
    \includegraphics[width=\linewidth]{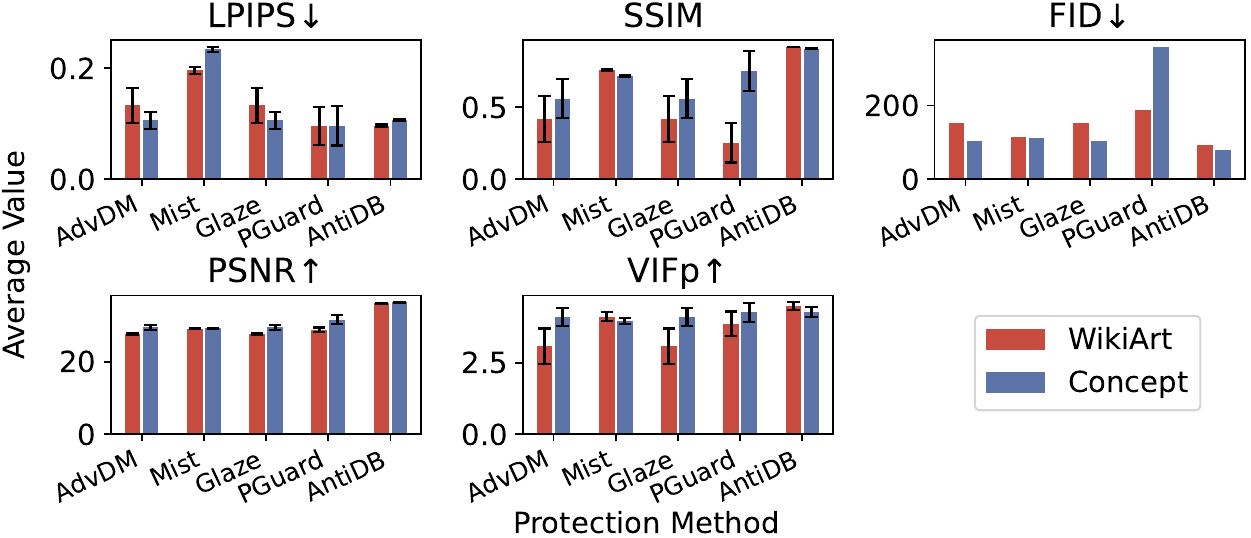}
    \caption{Fidelity evaluation of \textsc{Op}.} 
    \label{fig:fidelity_data}
\end{figure}

{\bf Fidelity} -- For a protected image, it is crucial that it appears visually identical to the original to maintain the image's utility. High fidelity indicates better preservation of artistic and semantic values. To evaluate the fidelity of various \textsc{Op} protections, we use several widely-used metrics, such as LPIPS, SSIM, PSNR, VIFp, and FID, as outlined in Table \ref{tab:watermarking_requirements}. 
\naen{Figure \ref{fig:fidelity_data} and \ref{fig:vis-data-pro} quantitatively and qualitatively show the fidelity evaluation of these methods compared with the protected image with the originals, respectively.}

Figure \ref{fig:fidelity_data} illustrates that 
fidelity varies among different \textsc{Op} protection methods. AntiDB shows the highest fidelity, with the lowest LPIPS averaging around 0.1 and FID around 80, along with the highest SSIM (exceeding 0.9), PSNR (above 36), and VIFp (around 4.4) across datasets.
AdvDM and Mist also exhibit relatively low FID averaging around 110, suggesting better preservation of image quality. 
\tbd{This is likely due to these methods incorporating DMs in perturbation optimization (\textit{cf}. Table \ref{tab:protection-summary}), enhancing fidelity compared to methods relying solely on image encoder.}

Inconsistencies arise when comparing metrics like LPIPS and FID. For instance, PGuard's low LPIPS of around 0.095 suggests a high visual similarity to the original images, but its high FID of over 180 suggests poor overall fidelity across the dataset. This disparity may stem from the different focuses of these metrics: LPIPS emphasizes semantic and perceptual similarities and visual details, while FID assesses how well the generated images align with the distribution of the original dataset, considering broader structural and statistical properties. Therefore, relying on a single metric can be misleading, highlighting the need for diverse metrics to comprehensively evaluate fidelity.


These findings align with the visualizations in Figure \ref{fig:vis-data-pro}, where the AntiDB-protected images closely resemble the originals, while those from AdvDM and Mist maintain high similarity with only slight noise. \tbd{Notably, all three methods DM into their optimization, contributing to their fidelity advantage.} In contrast, images protected by Glaze and PGuard show more noticeable alterations. \jiang{Specifically,} Glaze introduces subtle and unique distortions, while PGuard results in a stretched appearance compared to the original images.
Overall, while fidelity differs across \textsc{Op} methods, all successfully preserve key visual characteristics, ensuring utility without compromising the viewer's experience.



{\hidecontent{\jiang{no analysis above leads to the remark?}
\naen{analysis in the metrics in red before (detailed), and visualization again (short) to read smoothly.  }\tbd{checking all the remark...}}}
\autoremark{\textsc{Op} methods that use diffusion models during the perturbation optimization process tend to achieve higher fidelity compared to those that rely solely on image encoders.}

\begin{figure}[t]
    \centering
    \includegraphics[width=\linewidth]{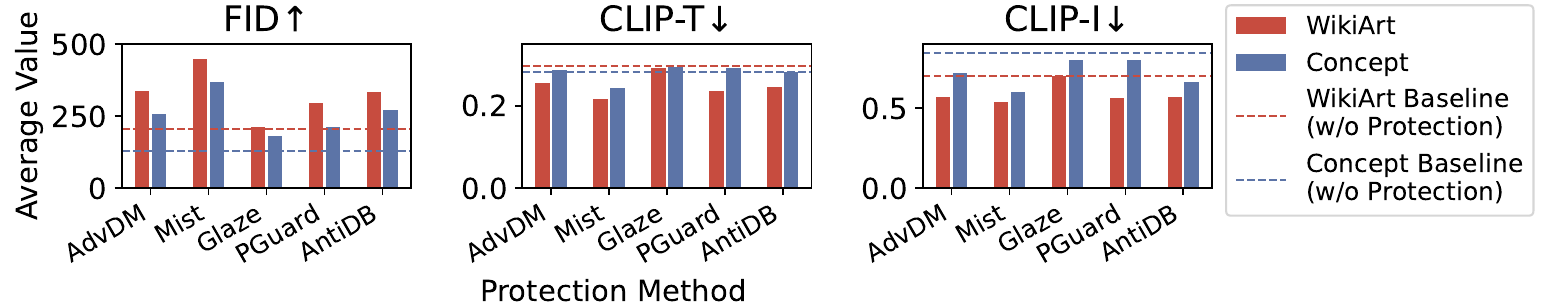}
    \caption{Efficacy evaluation of \textsc{Op}.}
    \label{fig:data_level_efficacy}
\end{figure}

{\bf Efficacy} -- \tbd{Following \cite{liang2023adversarial, shan2023glaze}, we fine-tune pre-trained SD models using protected images to generate mimicked images. For comparison, we also generate mimicked images from original (unprotected) images as a baseline.}
{\hidecontent{\jiang{original images?}\naen{$\checkmark$}}}
We assess the similarity between mimicked images produced from protected images and original images using FID, CLIP-I, and text-image alignment with CLIP-T. 
\tbd{Figure \ref{fig:data_level_efficacy} presents the quantitative results, and Figure \ref{fig:vis-data-pro} displays visual examples.} 




\begin{figure}[t]
    \centering
    \includegraphics[width=\linewidth]{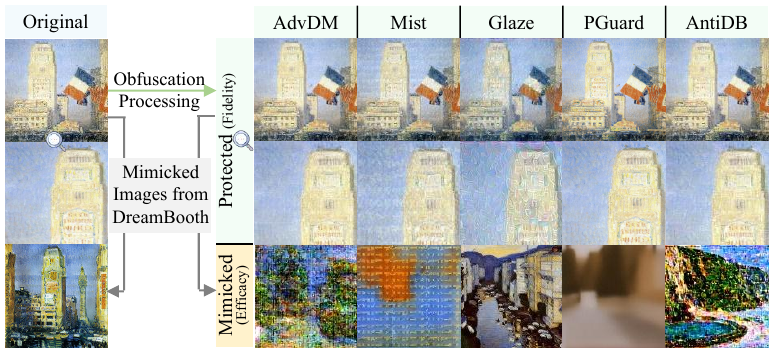}
    \caption{Fidelity and efficacy visualization of \textsc{Op} methods. Row 1\&3: protected images; row 2\&4: mimicked images.} 
    \label{fig:vis-data-pro}
\end{figure}


\tbd{As shown in Figure \ref{fig:data_level_efficacy}, mimicked images generated from protected images show a stronger deviation from originals, reflected in higher FID and lower CLIP-I and CLIP-T compared with originals, indicating efficacy in deterring copyright mimicry.}
Notably, Mist shows the highest efficacy, with an average FID increase (from around 150 to 400) and reductions in CLIP-I (from 0.7 to 0.55) and CLIP-T (from 0.30 to 0.23) across two datasets.
This indicates that mimicked images significantly diverge from the originals and their text descriptions, thus effectively mitigating copyright mimicry. 
Mist's strong efficacy \tbd{is due to} its incorporation of both image encoders and diffusion models into adversarial perturbation optimization\tbd{\cite{liang2023mist}}, effectively increasing latent space distance while minimizing pixel-level deviation. 
Other protections, such as AdvDM, AntiDB and PGuard, provide moderate protection with smaller changes in FID, CLIP-I, and CLIP-T, indicating subtler deviations.
In contrast, 
\jc{Glaze provides limited protection, as it slightly increases the FID (\textit{e.g.}, from 206 to 212 on the WikiArt dataset) while also slightly reducing both CLIP-T and CLIP-I. This result can be partly explained by the differences in fine-tuning methods used for image mimicry}, as DreamBooth differs from the fine-tuning methods employed in Glaze (details in Sec \ref{sec:generalizability}). 

{\hidecontent{\jiang{indicating ...}\naen{$\checkmark$}\jiang{speculation or conclusion?}\naen{$\checkmark$} \naen{move here to see the length of final line}}}
Figure \ref{fig:data_level_efficacy} demonstrates that the efficacy of protection methods varies across datasets. For instance, in the WikiArt dataset, almost all protection methods significantly reduce the similarity between generated images and text descriptions, as quantified by CLIP-T.
However, in the Concept dataset, only Mist shows a reduction in CLIP-T from 0.3 to 0.28, while other methods remain close to the baseline. 
This suggests that protecting Concept is more challenging than WikiArt, likely due to two factors:
(\textit{i}) many protection methods \cite{liang2023adversarial, liang2023mist, shan2023glaze} are optimized for artwork, enhancing their performance in art-centric datasets like WikiArt, and  
(\textit{ii}) 
the distinct styles in WikiArt are easier for protection methods to exploit, which
\jiang{underscores the complexity of evaluating protection methods across different datasets.} These findings highlight a need for more adaptable protection methods catering to varied data characteristics and contexts.

Notably, fidelity and efficacy do not always align. 
While stronger protections typically lead to greater quality degradation, our observations reveal counterintuitive results.
For instance, 
AntiDB exhibits strong fidelity (\textit{cf.} Figure \ref{fig:fidelity_data}) but does not achieve the best performance in efficacy (\textit{cf.} Figure \ref{fig:data_level_efficacy}). Similarly, Mist shows high efficacy but ranks moderately in fidelity. These showcase the complex interplay between fidelity (preserving image quality) and efficacy (ensuring robust copyright protection against mimicry), underscoring the need for a balance between visual quality and protection effectiveness in practical applications.

These quantitative findings align with visualization results in Figure \ref{fig:vis-data-pro}, where mimicked images from protected images show distinct styles from the originals.
Mist shows the most unique textures (highest efficacy), 
while AdvDM and AntiDB show artifacts (moderate efficacy). 
Recognizing the efficacy of each method is crucial for selecting optimal strategies to prevent mimicry of copyrighted content.

{\hidecontent{\jiang{why do you put figure 4 in before figure 5?}\tbd{fidelity metrics -- efficacy metrics -- fidelity and efficacy visualization}}}

\begin{figure}[t]
    \centering
    \includegraphics[width=\linewidth]{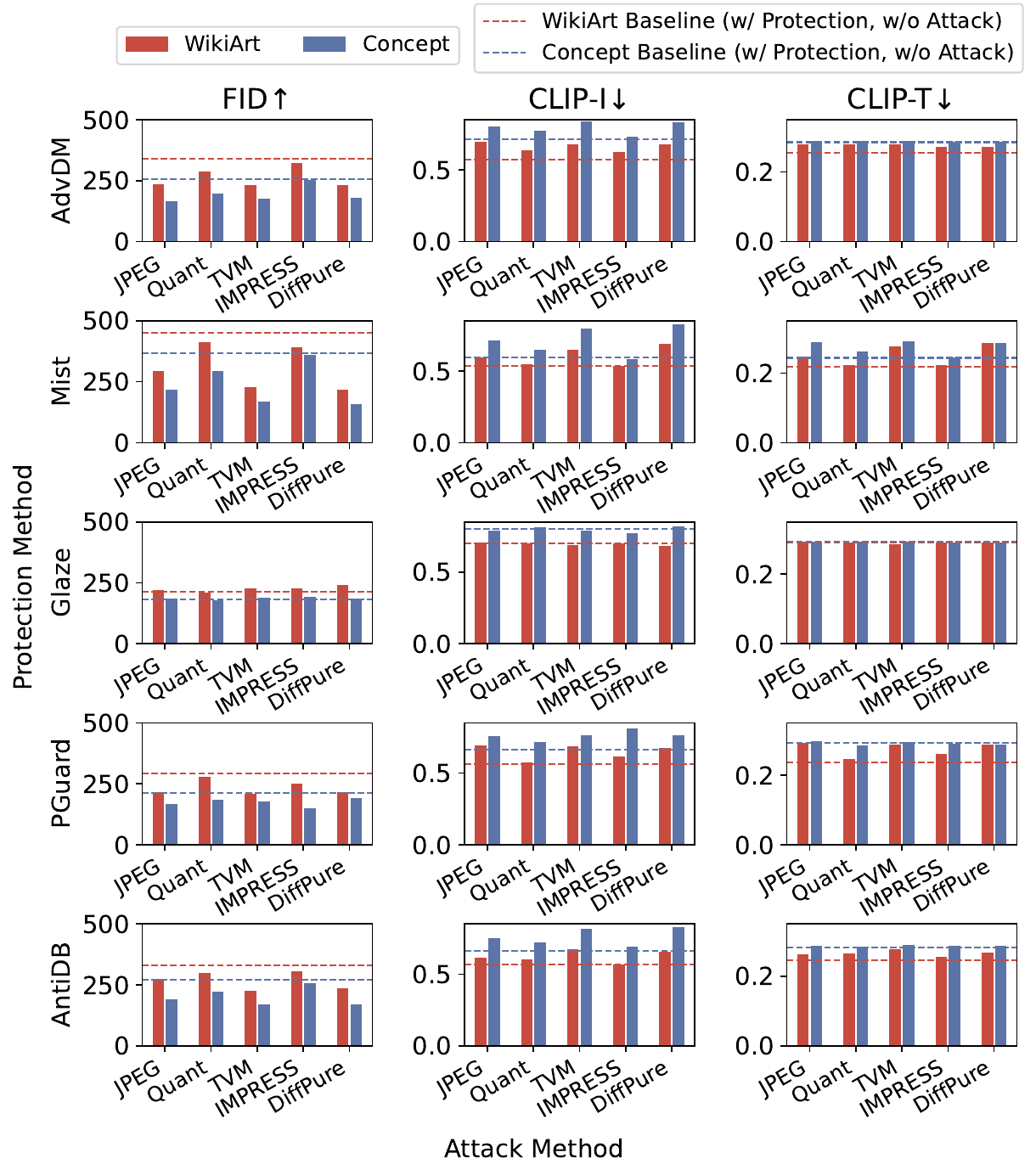}
    \caption{Resilience evaluation of \textsc{Op} against \textsc{Np}.}
    \label{fig:resilience_data}
\end{figure}


\begin{figure}[tp]
    \centering
    \includegraphics[width=\linewidth]{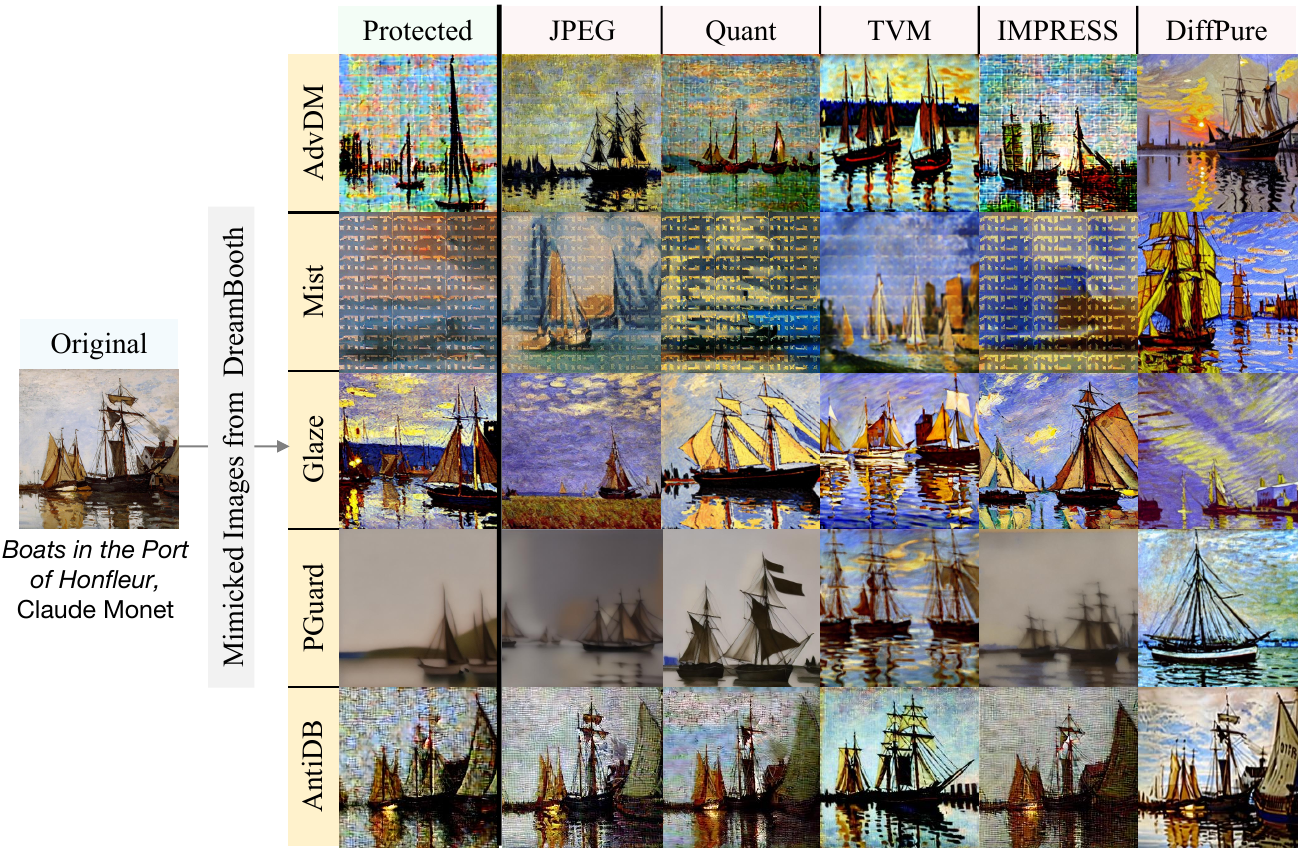}
    \caption{Resilience visualization of \textsc{Op} against \textsc{Np}. Column 1: mimicked images generated from protected images. Column 2-5: mimicked images generated from attacked images.} 
    
    \label{fig:vis-data-att-db-new}
\end{figure}

{\bf Resilience} -- \jiang{Following the approach outlined by \cite{cao2024impress}, we assess the resilience of \textsc{Op} protection methods against \textsc{Np} attacks. Our evaluation process involves fine-tuning Stable Diffusion (SD) models using purified images generated by applying \textsc{Np} to protected images (\textit{i.e.}, copyrighted images with \textsc{Op} applied). We then evaluate the mimicry performance of these fine-tuned models, where a higher mimicry performance indicates lower resilience of the protection method. Figure \ref{fig:resilience_data} presents the resilience evaluation results of various \textsc{Op} protection methods against \textsc{Np} attacks.}


\jiang{Analysis of Figure \ref{fig:resilience_data} reveals several key insights. 1) All protection methods, except Glaze, show a notable decline in effectiveness when subjected to purification attacks, as evidenced by higher mimicry performance. For instance, AdvDM-protected images, when purified, achieve lower FID and higher CLIP-I and CLIP-T compared to their unpurified counterparts, indicating a higher mimicry performance. Note that, Glaze’s apparent resilience stems more from its initially limited protection performance than superior defensive capabilities. 2) Different \textsc{Op} methods show varying protection abilities when applying attacks. For example, thanks to its initial strong protection performance, Mist still maintains relatively higher FID and lower CLIP-I and CLIP-T than other protection methods under attack. 3) TVM and DiffPure emerge as the most potent methods for diminishing \textsc{Op} protection, achieving higher mimicry performance. 4) CLIP-T shows less sensitivity than other metrics, especially on the Concept dataset where it remains nearly constant across most attack methods. We believe this is due to its robustness to minor protection artifacts, with significant changes only when major distortions obscure the original concept content.}

\jiang{Furthermore, we visualize the mimicry results of various \textsc{Np} methods against \textsc{Op} techniques in Figure \ref{fig:vis-data-att-db-new}. Our visual findings align with the quantitative analysis presented earlier. Specifically, 
Mist demonstrates superior protection performance even when NP methods are applied, with the exceptions of TVM and DiffPure. This observation further underscores that TVM and DiffPure are the most potent methods for diminishing \textsc{Op} protection: the artifacts in the mimicry images under TVM and DiffPure are notably less pronounced compared to other methods. Additionally, we observe that while \textsc{Np} can indeed diminish protection to some extent, certain \textsc{Op} protection methods still demonstrate a robust ability to prevent mimicry effectively. For instance, we can discern obvious protection patterns for Mist and AdvDM even after the application of \textsc{Np}.}

\jiang{In summary, both quantitative and qualitative analyses demonstrate that \textsc{Op} techniques can be compromised by certain attacks. These findings underscore the critical importance of evaluating protection methods not only for their initial effectiveness but also for their resilience against subsequent attacks. This comprehensive approach to assessment is essential for developing robust and reliable protection strategies in the face of evolving threats.}

\autoremark{TVM and DiffPure serve as dominant attacks to test the lower bounds of resilience in \textsc{Op} protection methods.}

\subsection{Model Sanitization Evaluation}
\label{sec:ms}
\label{sec:ms-evaluation}

Similar to Sec \ref{sec:op}, we assess model sanitization (\textsc{Ms}) across three key dimensions: \textit{fidelity}, \textit{efficacy}, and \textit{resilience}. Fidelity measures the sanitized model's ability to maintain performance on unrelated content. Efficacy gauges how effectively the sanitized model prevents the generation of copyrighted content, evaluating the thoroughness of the sanitization process. Resilience examines the sanitized model's robustness against concept recovery (\textsc{Cr}) attacks, assessing whether it consistently avoids reproducing copyright-protected concepts even under adversarial conditions. \tbd{The detailed experimental setup is given in Appendix \ref{app:protection-model}.} 

{\bf Fidelity} -- 
For sanitized models, it is crucial that sanitization preserves the ability to generate images for other concepts while excluding the copyright concept. Table \ref{tab:model-fidelity} evaluates the fidelity of \textsc{Ms} methods on MS-COCO 2017 \cite{lin2014microsoft} 30K dataset prompts. We use FID to measure the differences between images generated by the sanitized models (with the original DM for reference) and real-world images from the dataset. Additionally, CLIP-T is used to assess the alignment between generated images and prompts.


\jiang{Our analysis reveals that model sanitization (\textsc{Ms}) methods achieve a successful balance between copyright protection and image generation capabilities, with only minor impacts on overall performance. \naen{In Table \ref{tab:model-fidelity}, }sanitized models experience a marginal increase in FID scores compared to their original counterparts, with SLD showing the most notable change (16.95 vs. 16.21 for the original SD model). This subtle increase suggests a minor impact on image fidelity, likely due to the model inadvertently altering representations of unrelated but adjacent concepts or facing creative constraints when adjusted to exclude copyrighted content. Interestingly, CLIP-T scores remain remarkably consistent across all methods (0.30-0.31), indicating well-preserved textual alignment. These findings align with previous research \cite{gandikota2024unified,schramowski2023safe,gandikota2023erasing}, confirming that while \textsc{Ms} methods may slightly affect image fidelity, they successfully maintain text alignment for unrelated concepts. }

\jiang{In summary, the sanitization process achieves its primary goal of removing specific content without significantly compromising overall performance, demonstrating an effective balance between protecting copyrighted material and maintaining generative capabilities.}

\begin{table}[t]
\centering
{\footnotesize
    \centering
    \caption{Fidelity evaluation of \textsc{Ms}.}
    \renewcommand{\arraystretch}{0.8}
    \setlength{\tabcolsep}{4pt}
    \begin{tabular}{c|c||c|c|c|c|c|c}
    \textbf{Method} & \textbf{SD} & \textbf{FMN} & \textbf{ESD} & \textbf{AC} & \textbf{UCE} & \textbf{NP} & \textbf{SLD} \\ 
    \hline
    \hline
    \textbf{FID} $\downarrow$ & \textbf{16.21} & 16.47 & 16.51 & 16.95 & 16.64 & 16.89 & 16.95 \\ 
    \hline
    \textbf{CLIP-T} $\uparrow$ & \textbf{0.31} & 0.30 & 0.30 & 0.31 & 0.31 & 0.30 & 0.30 \\ 
    \end{tabular}
    \label{tab:model-fidelity}
}
\end{table}

\begin{figure}[t]
    \centering
    \includegraphics[width=\linewidth]{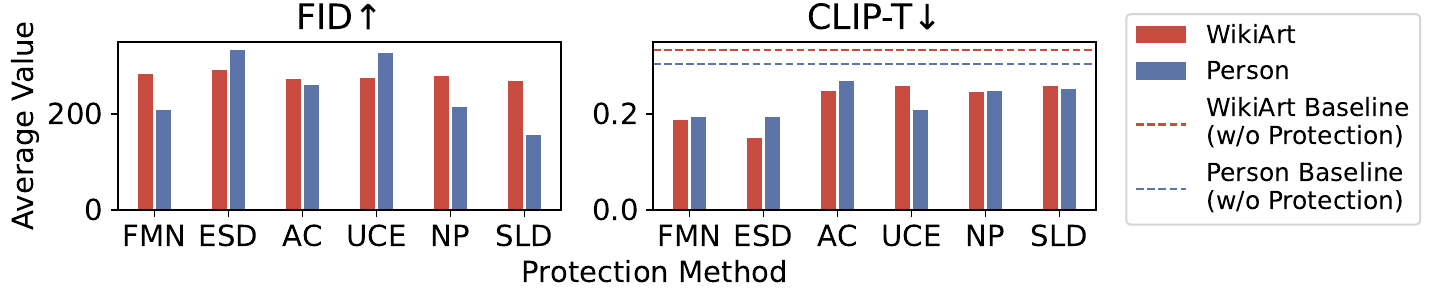}
    \caption{Efficacy evaluation of \textsc{Ms}.}
    
    \label{fig:efficacy_model}
\end{figure}

\begin{figure}[t]
    \centering
    \includegraphics[width=0.95\linewidth]{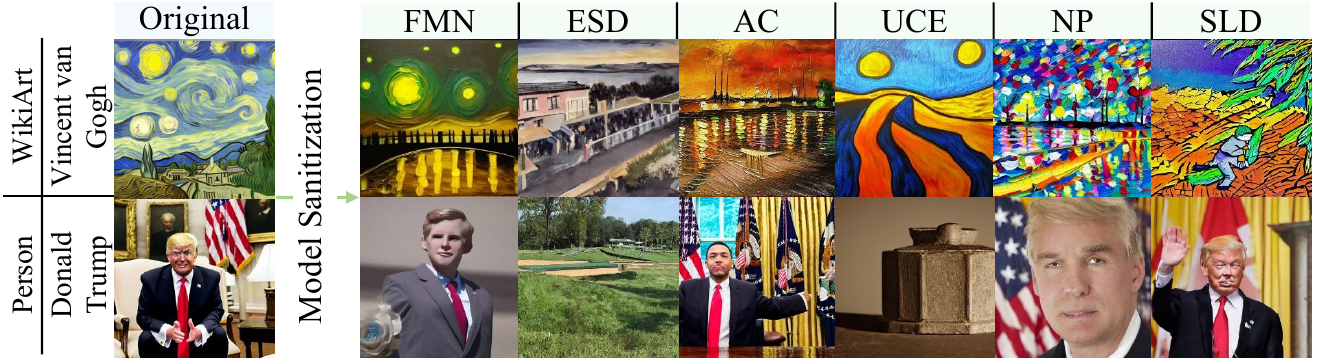}
    \caption{Efficacy visualization of \textsc{Ms}. Column 1: original images; column 2-7: sanitized models' outputs.} 
    \label{fig:vis-model-pro}
\end{figure}





{\bf Efficacy} -- \jiang{To evaluate the effectiveness of various model sanitization (\textsc{Ms}) methods in removing copyrighted concepts, we focus on two key metrics: image similarity and text-image alignment. We compare images generated by sanitized models to those from the original model using FID and measure the alignment between generated images and their prompts using CLIP-T scores. A higher FID or a lower CLIP-T implies a more effective \textsc{Ms} method.}

Figure \ref{fig:efficacy_model} reveals the variation in FID across images generated from different sanitized models. First, higher FID reflects more efficacy in removing copyright concepts from the sanitized model's outputs, while CLIP-T also shows a marked decrease from the baseline (original model alignment), suggesting great divergence from copyright content, with ESD performing best sanitization (average FID 311, CLIP-T 0.17). 
Notably, model fine-tuning methods (\textit{i.e.}, ESD, FMN, and UCE) generally outperform inference-guiding methods (\textit{i.e.}, NP and SLD) with higher FID and lower CLIP-T, reflecting more effective sanitization. \naen{This is likely because fine-tuning methods directly modify model parameters for deeper adjustments to reduce the retention of copyrighted content, while inference-guided methods only adjust output directions, resulting in superficial removals.}


Visualizations in Figure \ref{fig:vis-model-pro} support these findings. Fine-tuning-based methods like ESD and UCE effectively sanitize artistic styles by visibly altering original textures and colors in WikiArt and portraits into non-face images (\textit{i.e.}, landscapes or still lifes) in Person. In contrast, inference-guided methods like SLD still leave faint traces of original artistic style or individual characteristics.
Additionally, these categories differ significantly in time efficiency (\textit{cf}. Sec \ref{sec:efficiency}).



\autoremark{Fine-tuning-based \textsc{Ms} methods show greater efficacy than inference-guiding \textsc{Ms} methods.}

\begin{figure}[t]
    \centering
    \includegraphics[width=\linewidth]{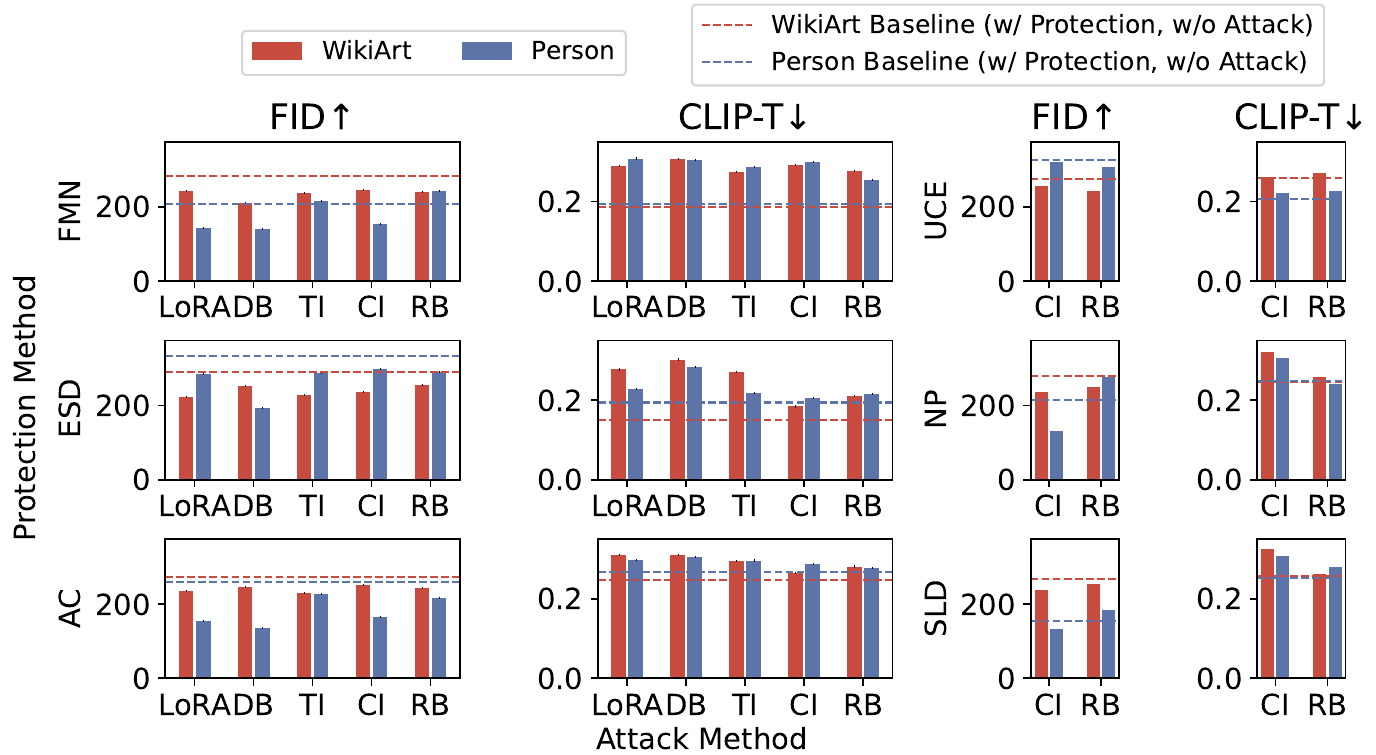}
    \caption{Resilience evaluation of \textsc{Ms} against \textsc{Cr}.} 
    \label{fig:resilence_model}
\end{figure}

{\bf Resilience} -- We evaluate the resilience of \textsc{Ms} against \textsc{Cr} attacks following \cite{pham2023circumventing}. Our evaluation process involves generating images from both original models (\textit{i.e.}, models capable of generating content with copyright concepts) and recovered models (\textit{i.e.}, sanitized models subjected to \textsc{Cr}). Figure \ref{fig:resilence_model} presents the resilience evaluation results of various \textsc{Ms} protection methods against \textsc{Cr} attacks.

\begin{figure}[t]
    \centering
    \includegraphics[width=\linewidth]{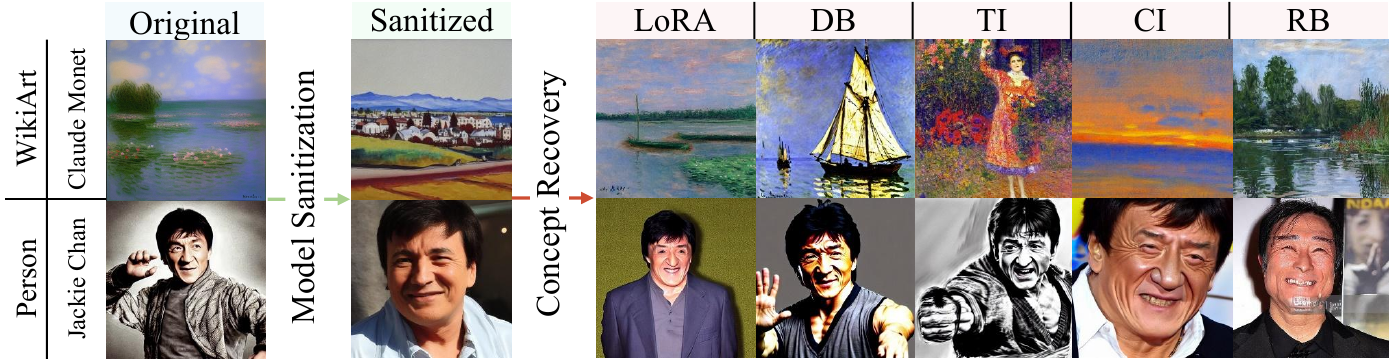}
    \caption{Resilience visualization of \textsc{Ms} against \textsc{Cr}. Column 1: original images; column 2: FMN-sanitized images; column 3-7: images generated from the recovered model.} 
    \label{fig:vis-model-att}
\end{figure}

Analysis of Figure \ref{fig:resilence_model} uncovers several critical insights. 
1) All \textsc{Ms} protection methods show reduced effectiveness under \textsc{Cr} attacks. The FID between images from recovered models and originals is lower than that of sanitized models and originals, with higher CLIP-T of images from recovered models than sanitized models, indicating enhanced resemblance to copyrighted content. 
For instance, 
FMN- and AC-sanitized models show relatively low resilience, with low FID scores and high CLIP-T under \textsc{Cr}. 
Thus, while \textsc{Ms} methods provide initial protection, their resilience against \textsc{Cr} attacks is limited. 
2) The resilience of \textsc{Ms} varies with \textsc{Cr} attacks applied. Fine-tuning-based attacks (\textit{e.g.}, LoRA and DB) are the most potent methods for diminishing \textsc{Ms} protection, lowering FID and raising CLIP-T from baseline. In contrast, textual-inversion-based attacks (\textit{e.g.}, TI and CI) cause moderate changes in FID and CLIP-T, while prompt-engineering-based attacks (\textit{e.g.}, RB) lead to minimal deviation. 
\naen{This may stem from incomplete pre-filtering of copyright content in DM's training dataset \cite{birhane2021multimodal,pham2023circumventing}, as \textsc{Ms} methods often remap them to new embeddings rather than fully remove these concepts.}
3) High-potency \textsc{Cr} attacks tend to limited in applicability. LoRA, DB, and TI are potent, but most apply to \textsc{Ms} using standardized open-source models. 
The custom CI pipeline for each \textsc{Ms} method makes it adjustable to various \textsc{Ms} methods, though its use with new \textsc{Ms} methods is uncertain. In contrast, RB bypasses protections solely through prompt modifications, making it adaptable across diverse T2I DMs. 

\tbd{Visualizations in Figure \ref{fig:vis-model-att} of images generated from the FMN-sanitized and recovered models reveal the varying resilience of \textsc{Ms} protection against \textsc{Cr} attacks with differing adversary capabilities. 
Attacks that enable deeper model manipulation, such as fine-tuning and textual-inversion methods, recover original styles in WikiArt and portrait characteristics in Person more effectively. This trend reflects a strong correlation between higher adversary capability and greater attack impact.
In contrast, less invasive prompt-engineering attacks have limited success in recovering detailed human portraits, but may still pose a feasible threat in scenarios with constrained adversary capabilities. These findings underscore the need for robust \textsc{Ms} methods that can withstand attacks across varying levels of attacker capability.
}

{\hidecontent{\jiang{todo}\naen{$\checkmark$}

\jiang{what is the relation between the remark and the above analysis? } \naen{$\checkmark$ add some transition in the last paragraph}}}

\autoremark{In real-world scenarios where models are accessible but not fine-tunable, the greatest threat to \textsc{Ms} methods comes from prompt engineering-based approaches.}

\subsection{Digital Watermarking Evaluation} 
\label{sec:dw}
\label{sec:dw-evaluation}


Similarly, we evaluate digital watermarking (\textsc{Dw}) based on three criteria: \textit{fidelity}, assessing the visual consistency between images before and after \textsc{Dw}; \textit{efficacy}, determined by the ACC of extracted watermark messages; and \textit{resilience}, \tbd{measuring the ACC of message extracted from image after watermark removal (\textsc{Wr}) attacks.} Further experimental setup details are outlined in Appendix \ref{app:protection-output}.

{\hidecontent{\jiang{todo}\naen{$\checkmark$}}}

\begin{figure*}[t]
    \centering
    \begin{minipage}{0.7\textwidth}
        \centering
        \includegraphics[height=2cm]{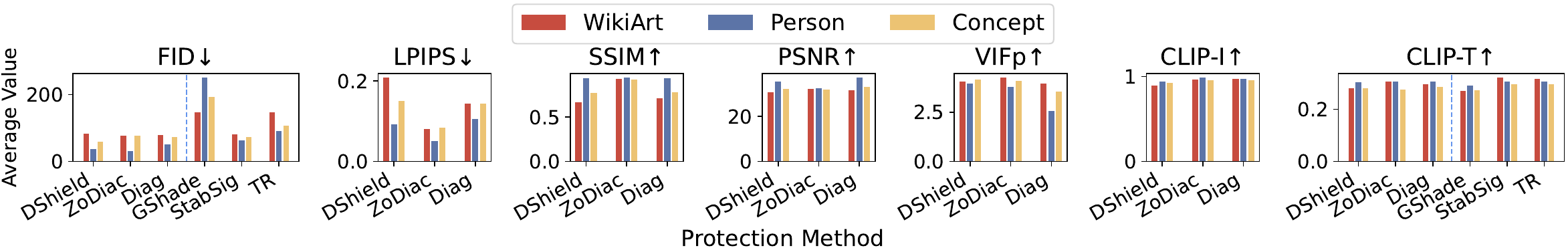}
        \caption{Fidelity evaluation of \textsc{Dw}.} 
        \label{fig:efficacy_output2}
    \end{minipage}%
    \begin{minipage}{0.3\textwidth} 
        \centering
        \includegraphics[height=2cm]{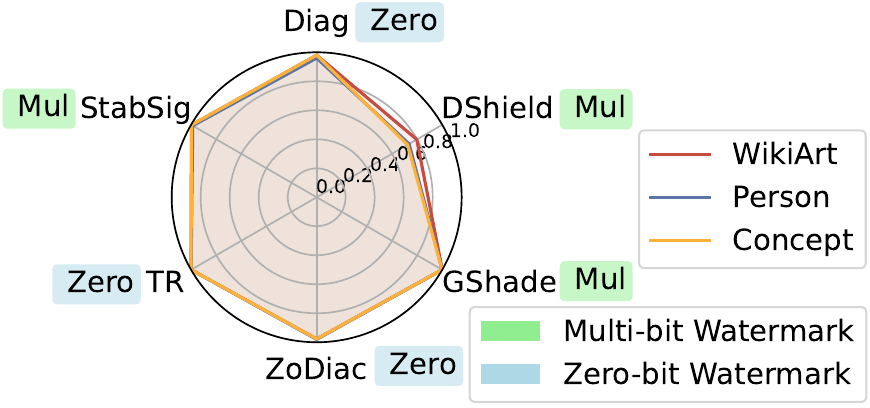}
        \caption{Efficacy evaluation of \textsc{Dw}.}
        \label{fig:acc_values_transposed}
    \end{minipage}
\end{figure*}


\begin{figure}[t]
    \centering
    \includegraphics[width=\linewidth]{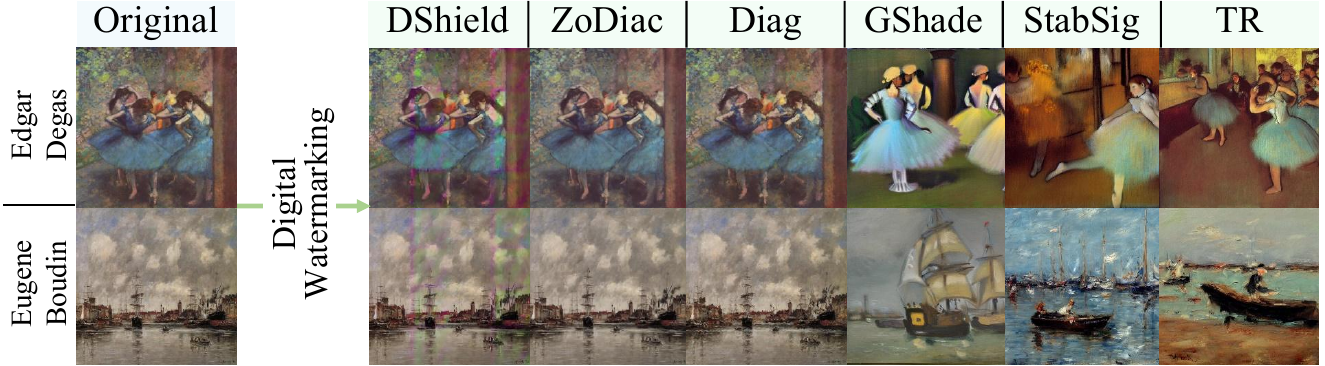}
    \caption{Fidelity visualization of \textsc{Dw}. Column 1: original images; column 2-7: watermarked images.}
    \label{fig:vis-output-pro}
\end{figure}


{\bf Fidelity} -- Maintaining visual similarity to the original image and alignment with the corresponding prompt is essential for watermarked images. 
\tbd{Figure \ref{fig:efficacy_output2} evaluates fidelity across all \textsc{Dw} methods, using FID as a general metric. Specifically, for watermarks embedded directly onto existing images, we measure visual consistency with metrics such as LPIPS and SSIM; for generative watermarks that produce watermarked images from prompts, we assess text alignment with CLIP-T.}
Visualizations are presented in Figure \ref{fig:vis-output-pro}.


{\hidecontent{\jiang{Figure 12}\naen{$\checkmark$}}}

Figure \ref{fig:efficacy_output2} shows that \textsc{Dw} have minimal impact on image fidelity, where lower LPIPS and higher SSIM, PSNR, VIFp, and CLIP-I suggest greater fidelity. Specifically, DShield, ZoDiac, and Diag exhibit low FID (below 80), indicating minimal visual alteration. 
\tbd{This is attributed to its approach of embedding the watermark in the latent space's Fourier frequencies, making disturbances less visually perceptible \cite{zhang2024robust}.
In contrast, GShade, StabSig, and TR display slightly higher FID (exceeding 90) but maintain CLIP-T scores comparable to watermarks on existing images (around 0.3), indicating preserved semantic consistency.}


{\hidecontent{\jiang{existing images?}\naen{$\checkmark$}
\jiang{metric}\naen{$\checkmark$}
\jiang{reason}\naen{$\checkmark$}

\jiang{if  these methods are designed for watermarking existing images, why do you use them to evaluate other metrics?} \naen{$\checkmark$}}} 

Visualizations in Figure \ref{fig:vis-output-pro} confirm these findings. DShield, ZoDiac, and Diag retain a close resemblance to the originals, while GShade, StabSig, and TR introduce differences, with content and artistic style largely unchanged at the semantic level. This consistency across metrics and visuals supports the fidelity of these watermark designs.

{\hidecontent{\jiang{todo}}}

{\bf Efficacy} --  For watermarks, it is crucial that the decoded message exhibits high ACC compared to the embedded message. High efficacy indicates better copyright verification. 

Figure \ref{fig:acc_values_transposed} shows that most \textsc{Dw} methods achieve ACC close to 100\% across datasets, except for DShield, underscoring the efficacy of these watermarks.
Notably, TR stands out with 100\% ACC across all datasets, indicating robust watermark embedding and decoding capability. 
\tbd{DShield shows slightly lower ACC, possibly due to the diverse and complex datasets we used, 
underscoring a limitation of fine-tuning-based watermarking methods, where efficacy depends on data specificity and quality.}

\tbd{In summary, these findings highlight that the efficacy of \textsc{Dw} is largely dependent on the embedding strategy or the quality of training data, while modifications in latent space show particular promise for high ACC in diverse settings.}


{\hidecontent{\jiang{last sentence?}\naen{$\checkmark$}}}

\begin{figure}[tp]
     \centering
    \includegraphics[width=\linewidth]{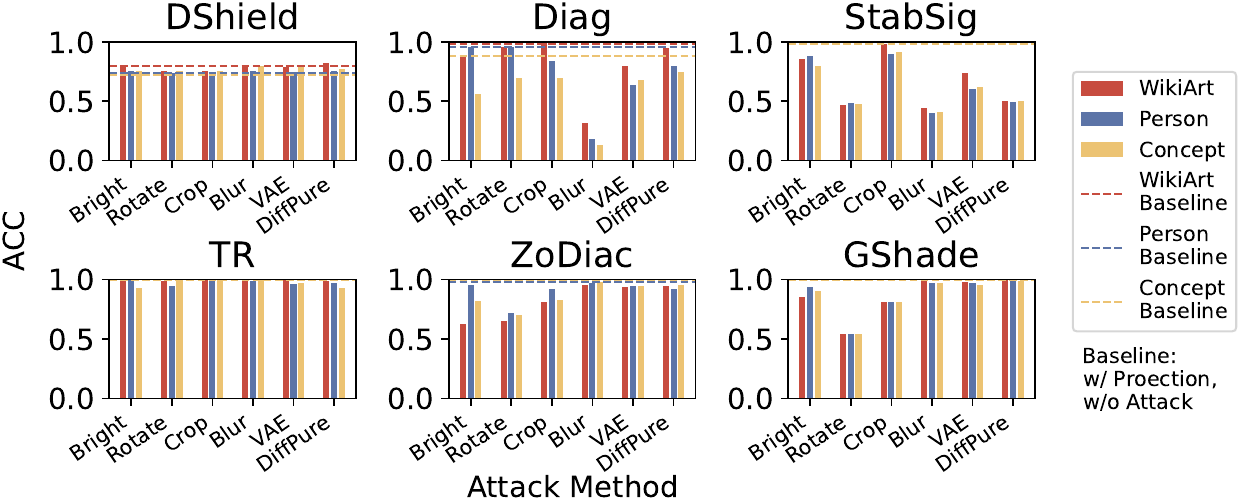}
    \caption{Resilience evaluation of \textsc{Dw} against \textsc{Wr}. }
    \label{fig:resilence_output}
\end{figure}

{\bf Resilience} --  We assess \textsc{Dw} protection resilience against \textsc{Wr} attacks by comparing the ACC of messages extracted after watermark removal with the originally embedded message. A higher ACC indicates a stronger resilience.


Figure \ref{fig:resilence_output} presents the resilience of various \textsc{Dw} protections against \textsc{Wr} attacks, revealing several key insights. 
1) Most watermarks show reduced protection after attacks, with ACC lower than the baseline (un-attacked watermarked images). For example, Diag's ACC sharply declines under Blur, while StabSig, ZoDiac, and GShade are vulnerable to Rotate. 
2) \textsc{Dw} methods vary in resilience against attacks. Compared to the baseline, DShield and TR exhibit only slight declines under attacks, while others face larger reductions under certain attacks. 
3) Under attack, latent space modifying methods exhibit higher ACC compared to model fine-tuning methods, with TR maintaining nearly 100\% ACC across attacks due to its invisible Fourier space embedding that resists pixel disruption. 
4) ZoDiac and GShade share similar vulnerabilities under Bright, Rotate, and Crop attacks, with the lowest ACC observed under Rotate.

In summary, these insights highlight the need to carefully consider specific attack scenarios when choosing watermark strategies. 
We speculate that latent-space modifying methods leverage the inherent distribution of the diffusion model's latent space to embed watermarks more subtly and securely, making them harder to detect and remove. 
\autoremark{Latent space-based watermarks tend to be more resilient against attacks than fine-tuning-based watermarks.}


\section{Exploration}


\naen{Next, we explore the generalizability, efficiency, and sensitivity of current protection methods. We further compare these methods with their contemporary versions and industry-leading online text-to-image applications.
Furthermore, we also conduct user studies to evaluate the alignment between evaluation metrics and human judgment.}


\begin{figure}[t]
    \centering
    \begin{minipage}{0.62\linewidth}
        \centering
        \includegraphics[height=2.9cm]{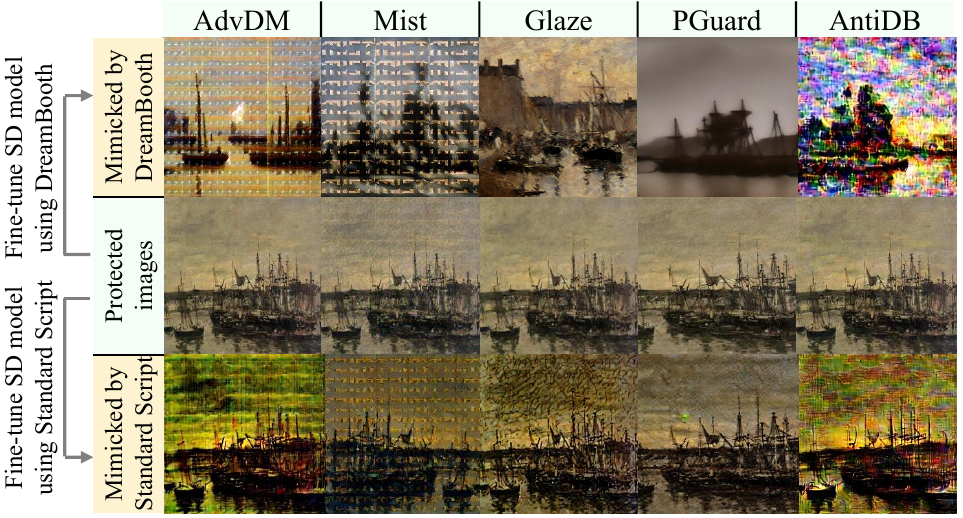}
        \caption{Various style mimicry methods based on fine-tuning. } 
        \label{fig:Finetuning}
    \end{minipage}%
    \hspace{0.01\linewidth}
    \begin{minipage}{0.35\linewidth} 
        \includegraphics[height=2.9cm]{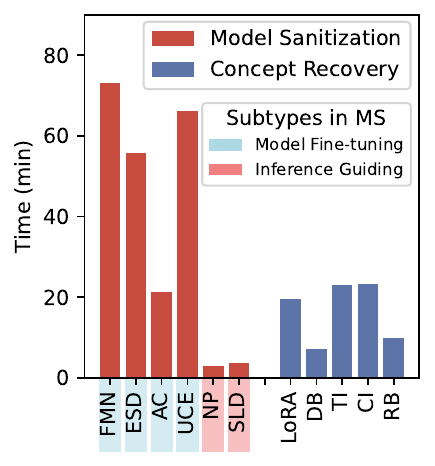}
         \centering
        \caption{Efficiceny evaluation.}
        \label{fig:time-efficiency}
    \end{minipage}
\end{figure}


\subsection{Generalizability}
\label{sec:generalizability}



While previous experiments use DreamBooth\cite{ruiz2023dreambooth} for image mimicry, other fine-tuning methods can also achieve mimicry. To assess generalizability, \minxi{following \cite{cuidiffusionshield,honig2024adversarial}, we employ a standard script from Diffusers\footnote{\url{https://huggingface.co/docs/diffusers/training/text2image}} for fine-tuning (details in Appendix \ref{app:minicry-dreambooth}).} 
In Figure \ref{fig:Finetuning}, differences in texture patterns, artifacts, or deviations from protected images reveal protection effectiveness against specific mimicry techniques. Notably, AdvDM and Mist exhibit the highest generalizability, providing strong protection under both fine-tuning methods. Glaze is less effective against DreamBooth but performs better with the standard script. Conversely, PGuard is robust with DreamBooth but is less effective with the standard script. AntiDB provides noticeable protection against both methods, particularly performing well against DreamBooth mimicry. These findings emphasize the need for protection methods that account for diverse mimicry techniques to enhance generalizability. 

\autoremark{The generalizability of various copyright protection methods differs significantly.}

\begin{figure}[t]
    \centering
    \includegraphics[width=\linewidth]{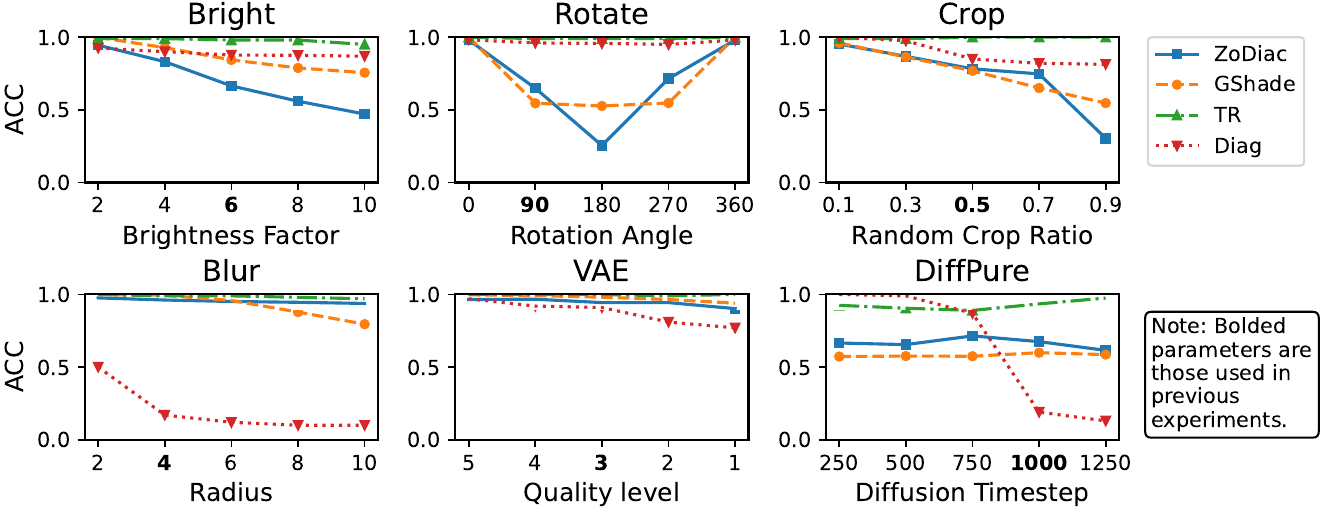}
    \caption{Sensitivity analysis on watermark removal.} 
    \label{fig:output_attack_par}
\end{figure}

\subsection{Efficiency} 
\label{sec:efficiency}



\naen{Computational cost is a key factor in copyright protection applications. \textsc{Op} and \textsc{Dw} involve lightweight, image-level manipulations, 
while \textsc{Ms} and \textsc{Cr} require deeper model-level changes, increasing time consumption. While prior studies often overlook time efficiency, we explore the time consumption of \textsc{Ms} and \textsc{Cr} methods.}

As shown in Figure \ref{fig:time-efficiency}, within \textsc{Ms}, inference-guiding methods are more efficient than model fine-tuning methods as they can sanitize a concept within four minutes without retraining. 
Additionally, \textsc{Cr} methods are generally more time-efficient than \textsc{Ms} methods on average. 
This is likely because \textsc{Cr} simply enhances existing representations in the model, whereas \textsc{Ms} must first overcome the model's training biases with a reverse optimization process.
Notably, DB is the most efficient \textsc{Cr} method, highlighting the vulnerability of \textsc{Ms}.
Therefore, practitioners should carefully select \textsc{Ms} methods based on available computational resources.

{\hidecontent{\jiang{why only Ms  and Cr} \naen{$\checkmark$}}}


\autoremark{In scenarios where efficiency is a priority, inference-guiding methods are preferred to model fine-tuning methods.
}

\subsection{Sensitivity Analysis}
\label{sec:sensitivity-analysis}





\tbd{Following \cite{wen2023tree,zhang2019robust,yang2024gaussian}, we take \textsc{Dw} as a representative for sensitivity analysis, where small perturbations may sharply impact watermark resilience, offering generalizable insights to other protection methods.}

{\hidecontent{\jiang{not that convincing}\naen{$\checkmark$}}}
As shown in Figure \ref{fig:output_attack_par}, most protection methods exhibit a sharp ACC drop with increased hyperparameter values. For instance, higher brightness, crop ratios, or blur radii result in ACC drops. This implies that strong hyperparameter settings weaken the robustness of most methods. 
Notably, both TR and Diag demonstrate notable resilience to the Rotate attack, maintaining near 100\% ACC even at 90$^{\circ}$ or 180$^{\circ}$ rotation, while GShade and ZoDiac suffer sharp decreases. The superior performance of TR is attributed to its multi-ring pattern in Fourier space. Similarly, Diag achieves robustness by embedding triggers to embed robust watermark patterns. These insights help practitioners choose protection methods tailored to real-world attack scenarios.

{\hidecontent{\jiang{only watermark? why} \naen{$\checkmark$}}}
\subsection{Contemporary Assessment} 

In the evolving field of copyright protection, methods and infringement attacks are in constant competition. 
We analyze recent versions of protections and attacks:
\naen{(\textit{i}) \textbf{Glaze v2.1\footnote{\url{https://glaze.cs.uchicago.edu/}}} is a closed-source update optimized for styles with clear colors and smooth textures. (\textit{ii}) \textbf{Mist v2\footnote{\url{https://psyker-team.github.io/}}} enhances the vanilla Mist \cite{liang2023mist} with improved efficacy and efficiency. (\textit{iii}) \textbf{Noisy Upscaler\cite{honig2024adversarial}} is an advanced attack that first adds a small amount of random noise to a protected image, then purifies the image using the Upscaler\cite{rombach2022high}. }

\naen{Figure \ref{fig:glaze2mist2} compares Glaze v2.1 (details in Appendix \ref{app:correctness-analysis}) with our open-sourced Glaze implementation, along with Mist v2 and vanilla Mist. Although both Glaze versions introduce similar perturbations, Glaze v2.1 still demonstrates limited resilience, especially against JPEG attacks. Mist v2 achieves improved resilience with post-attack images displaying noticeable mottling and color shifts, while images protected by the original Mist method show a closer resemblance to the originals. These observations underscore the vulnerabilities of existing copyright protection methods to advanced attacks, highlighting the ongoing need for improved protective solutions.
In Figure \ref{fig:mist+upscale}, current protections are particularly susceptible to the Noisy Upscaler attack, with heightened vulnerability compared to other methods. }

{\hidecontent{\jiang{hard to follow}\naen{$\checkmark$}}}

\autoremark{The arms race between protections and attacks promotes the development of more advanced protections.}

\begin{figure}[t]
    \centering
    \includegraphics[width=\linewidth]{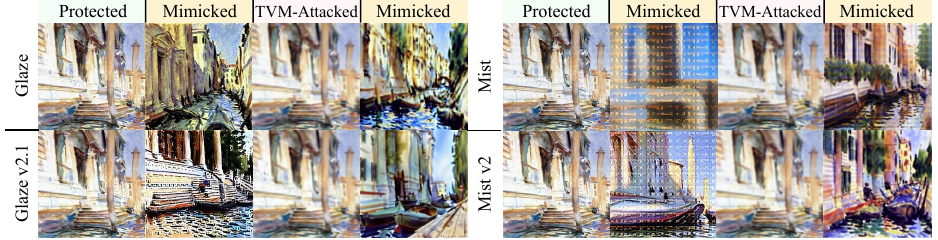}
    \centering
    \caption{Comparison of Glaze v2.1 and our open-sourced Glaze implementation, along with Mist v2 and Mist.}
    \label{fig:glaze2mist2}
\end{figure}

\begin{figure}[t]
   
    \includegraphics[width=\linewidth]{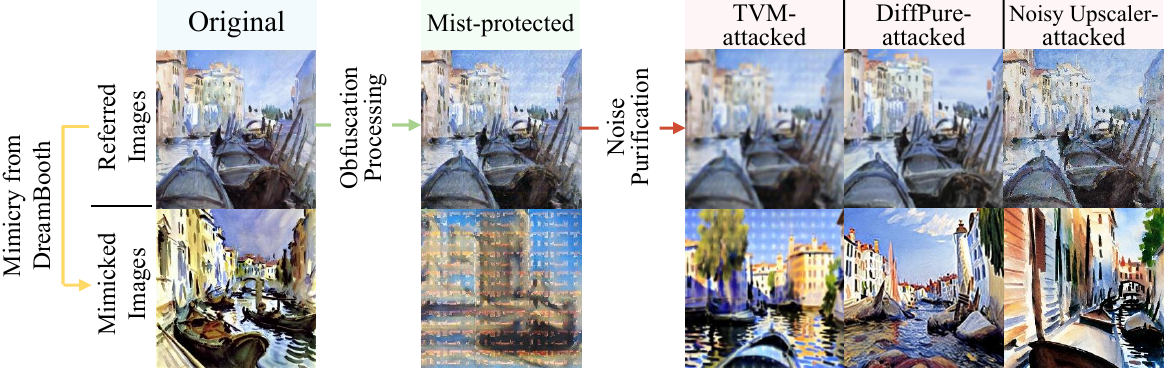}
     \centering
    \caption{The result of Mist and Noisy Upscaler. } 
    \label{fig:mist+upscale}
\end{figure}

\subsection{Real-world Online Applications}
\label{sec:realworld}


After analyzing SOTA strategies in academic settings, we compare them with industry-leading online applications. We have reported our findings to the respective companies.

\begin{figure}[t]
    \includegraphics[width=\linewidth]{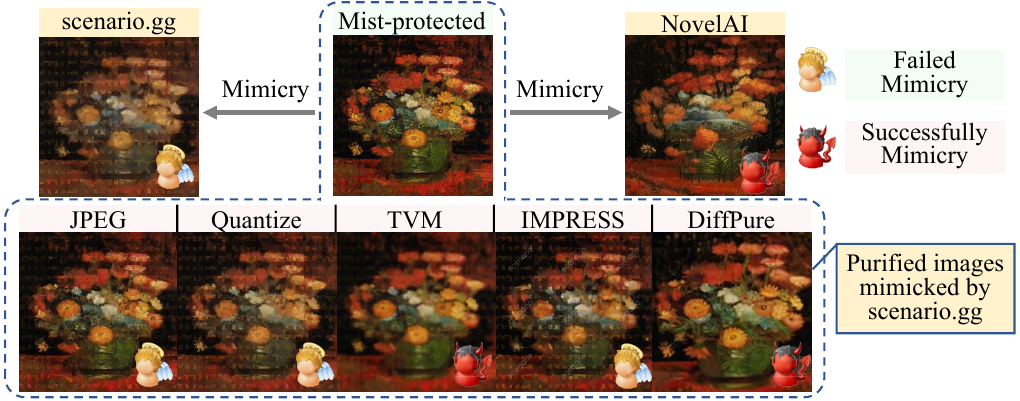}
     \centering
       \caption{Visualization results of scenario.gg and NovelAI.} 
    \label{fig:scenario-gg}
\end{figure}

\textbf{Scenario.gg and NovelAI for image mimicry.}
To assess the efficacy of \textsc{Op}, we explore two online applications, scenario.gg\footnote{\url{https://www.scenario.gg/}} and NovelAI\footnote{\url{https://novelai.net/}}.
Figure \ref{fig:scenario-gg} illustrates that Mist, the strongest protection, effectively prevents mimicry on scenario.gg, as the perturbation remains intact. 
Further, we observe that the mimicked images from TVM- and DiffPure-purified images make artifacts nearly undetectable, emerging as the most potent attacks, aligning with previous findings in Section \ref{sec:op}.
On NovelAI, its style transfer removes Mist's perturbation, suggesting that frequent model updates may reduce protection efficacy. This highlights the importance of ongoing protection updates to counter mimicry threats.




\begin{figure}[t]
    \centering
    \includegraphics[width=\linewidth]{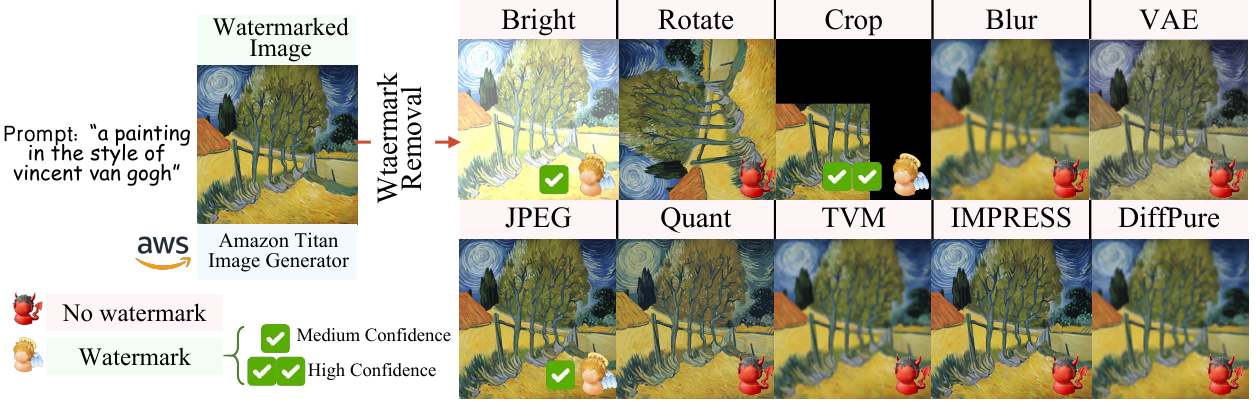}
    \caption{Watermark from Amazon Titan Image Generator.}
    \label{fig:amazon}
\end{figure}

\textbf{Amazon Titan Image Generator for watermarking.}
We assess the resilience of the watermark embedded in Amazon Bedrock Titan Image Generator\footnote{\url{https://aws.amazon.com/cn/bedrock/titan/}} against various attacks. 
Figure \ref{fig:amazon} shows the watermark remains intact against basic attacks (\textit{e.g.}, Bright, Crop, and JPEG), but becomes undetectable under more complex attacks. This implies that online watermarks share similar vulnerabilities to those applied locally. Additionally, \wenjie{online watermarks lack the customization options (\textit{e.g.,} watermark strength)}. Therefore, practitioners should take flexibility and customization into consideration when choosing watermark methods.


\subsection{User Study} 

We conduct user studies to evaluate the alignment between metrics and human perception (details in Appendix \ref{app:appendix-user-study}). 
First, we assess the visual quality and style mimicry of protected and purified images in WikiArt. Following \cite{honig2024adversarial}, we define success rate as the percentage of users preferring mimicry images fine-tuned on protected or purified images over unprotected ones. 
We observe that success rates increase after purification, suggesting greater visual similarity to the originals. Notably, average success rates across all mimicry scenarios remain below 50\% (50\% suggests perfect mimicry), showing that from human perspectives, even mimicked images from purified images still differ significantly from the originals. Mist yields the lowest mimicry success rate (under 10\%), indicating the highest efficacy for protection, whereas DiffPure attacks reduce resilience, with success rates around 35\%, supporting observations in Sec \ref{sec:op}. Second, following \cite{gandikota2023erasing,gandikota2024unified,schramowski2023safe}, we further examine whether \textsc{Ms} methods impact the fidelity of images of unrelated concepts. Over 50\% of users rate the fidelity and text alignment of images generated by sanitized models as equal to or better than the original SD model, supporting the observations in Sec \ref{sec:ms}, suggesting that sanitized models produce images comparable to those from the original SD. The alignment between metrics and human judgment confirms that \lib effectively captures human perception for assessing copyright protection methods.


{\hidecontent{\jiang{result?}\naen{$\checkmark$}}}

\section{Discussion}
\textbf{Limitations and future work.} First, \lib integrates most mainstream copyright protection methods in T2I DMs. Although it does not implement all strategies, its modular design allows easy incorporation of new protections, attacks, and metrics. Second, we primarily apply default settings from original papers, as these are typically optimized for performance. However, our framework supports alternative configurations. Finally, most protections require modifying original artworks, posing challenges for established artists whose unprotected works remain vulnerable. Unlike software security, where updates can fix vulnerabilities, copyright protection cannot be easily patched. 
As offense-defense dynamics evolve,  existing protections may not withstand future attacks. We hope that \lib provides interim protection and advocates for the establishment of more comprehensive laws and regulations.



\textbf{Guidance for enhancing protection methods.} Our findings reveal limitations in current copyright protections, with \lib serving as a valuable benchmark for improvement.  For example, adversarial perturbations in \textsc{Op} are easily compromised by simple attacks like JPEG, so incorporating JPEG loss into optimization may improve resilience. 
In \textsc{Ms}, resilience can be improved through adversarial training with crafted adversarial inputs that induce the generation of copyright concepts, minimizing model output probability under these inputs to facilitate concept erasure in more complex scenarios. Alternatively, if \textsc{Cr} is inevitable, refining \textsc{Ms} to slow recovery efforts provides additional protection. For \textsc{Dw}, designing watermarks with common attack strategies can strengthen resilience.

\textbf{Additional related work.} Recent studies \cite{ren2024copyright,pham2023circumventing,honig2024adversarial} have surveyed copyright protections and attacks methods in T2I DMs but are limited to single-level implementations without empirical evaluation. For instance, \cite{ren2024copyright} discusses \textsc{Op} and \textsc{Ms} without experimental validation or quality assessment of generated images. Similarly, \cite{honig2024adversarial} highlights \textsc{Op} artistic style imitation, showing that all existing copyright protections can be bypassed through user studies, but lack quantitative metrics for image quality. 
In contrast, \lib provides a comprehensive framework for evaluation, covering major protection and attack categories within a unified platform for empirical analysis.




\vspace{-0.15cm}
\section{Conclusion}
\vspace{-0.15cm}
In this paper, we design and implement \lib, a uniform platform dedicated to the comprehensive evaluation of copyright protection for text-to-image diffusion models.
Leveraging \lib, we conduct systematic evaluations from the perspectives of fidelity, efficacy, and resilience.
To our knowledge, this platform is the first of its kind to provide a uniform, comprehensive, informative, and extensible evaluation of existing copyright protections and attacks. It offers empirical support and addresses the under-explored intricacies of copyright protections and attacks that have previously suffered from non-holistic and non-standardized evaluations, thereby tackling long-standing questions in the field. 

\newpage
\bibliographystyle{IEEEtran}
\bibliography{bibs/reference}

\newpage
\appendices

\begin{figure}[!t]
    \centering
    \includegraphics[width=85mm]{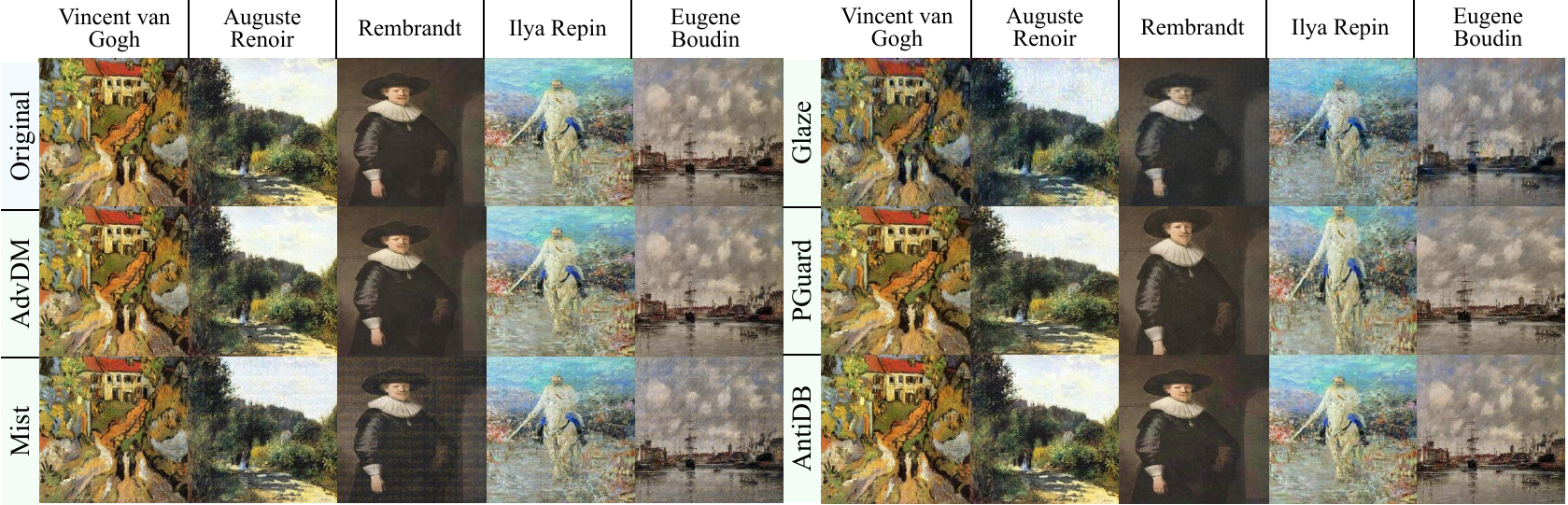}
    \caption{Fidelity visualization of \textsc{Op}. Column 1: original images; column 2-6: protected images.}
    \label{fig:data-level-protection}
\end{figure}

\begin{figure}[!t]
    \centering
    \includegraphics[width=85mm]{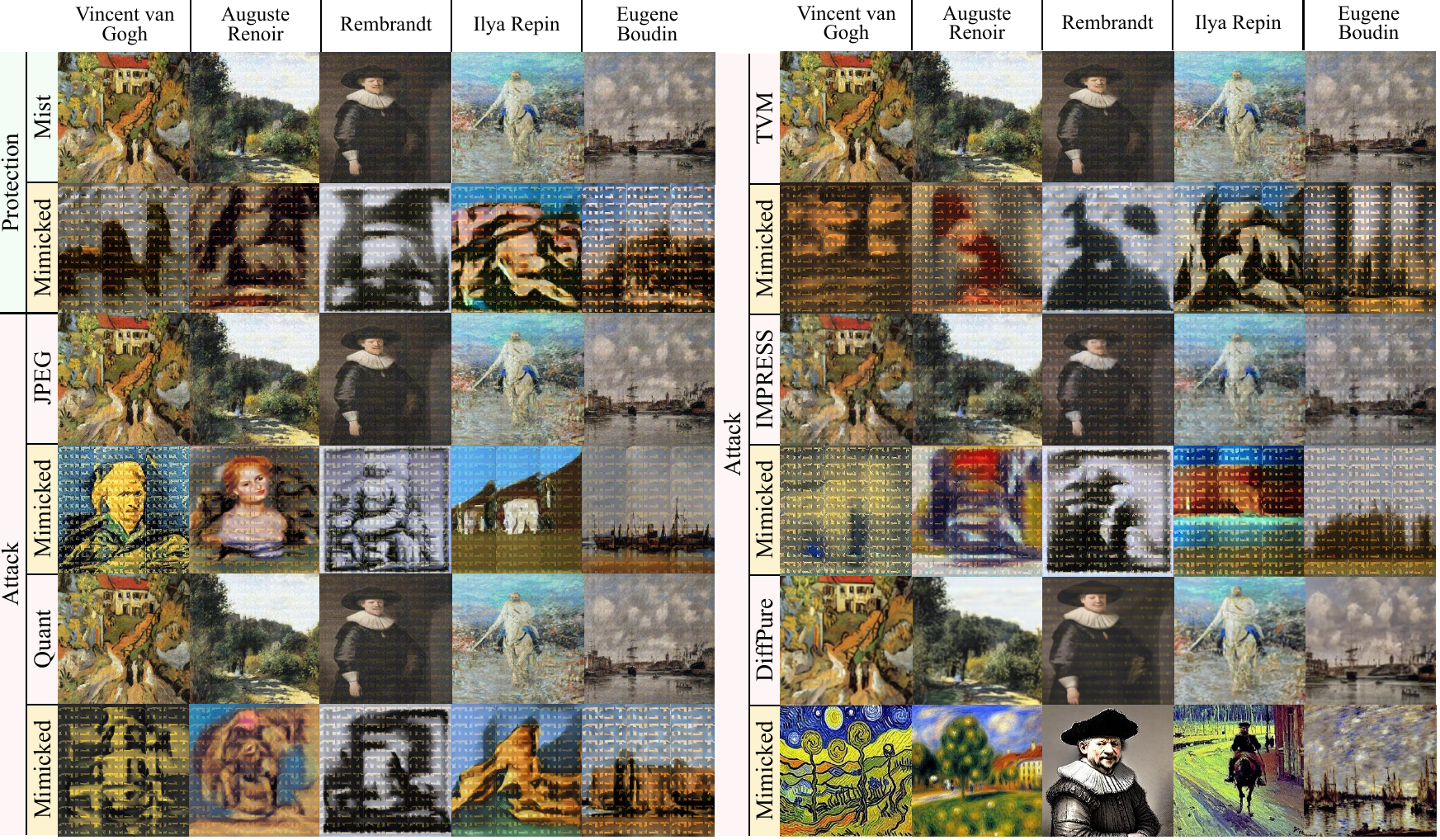}
    \caption{Efficacy visualization of \textsc{Op}. Row 1-2: Mist-protected image and its mimicked artwork; row 3-12: protected image under attacks and their mimicked artworks.}
    \label{fig:data-level-dreambooth-generation}
\end{figure}

\begin{figure}[!t]
    \centering
    \includegraphics[width=85mm]{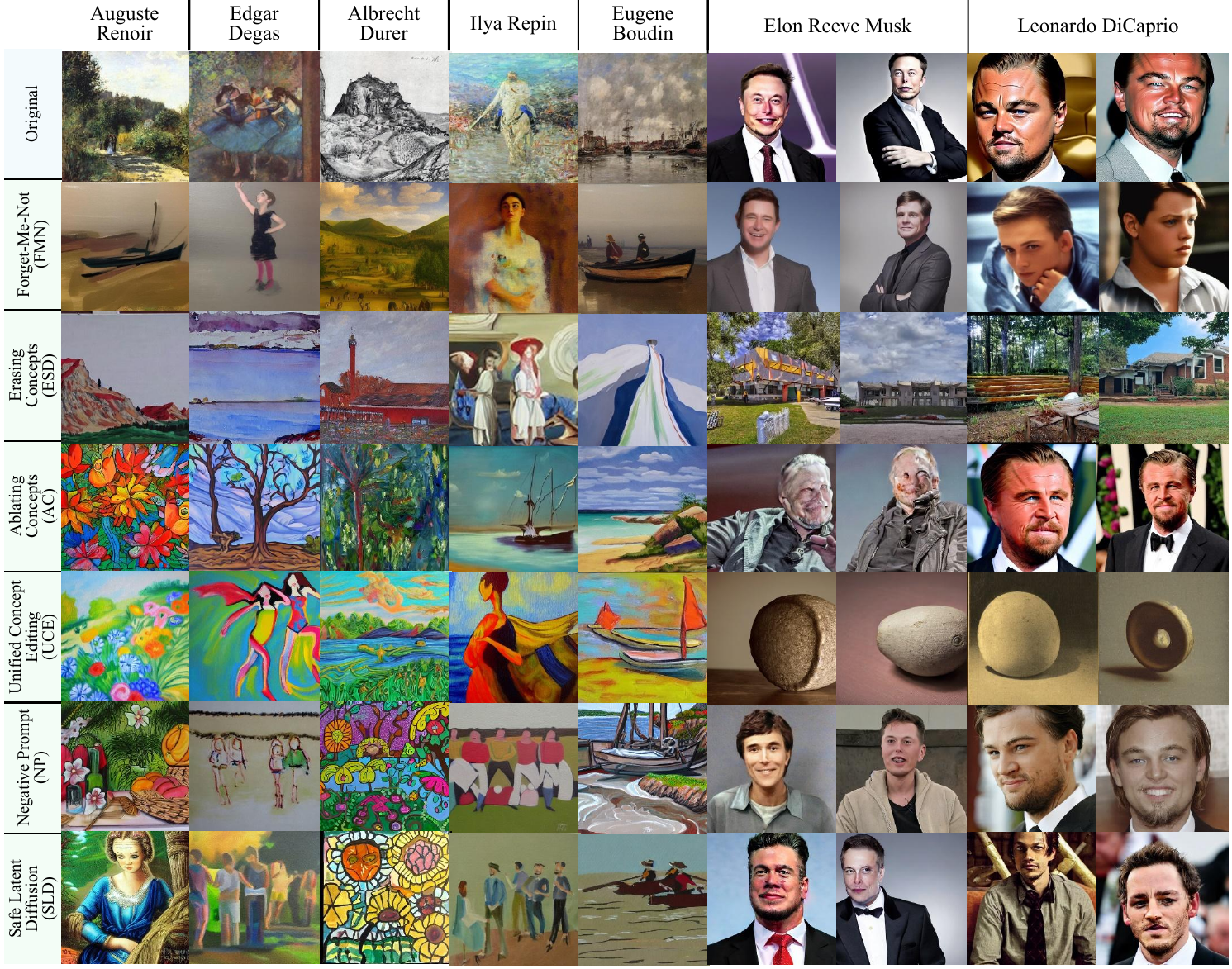}
    \caption{Efficacy visualization of \textsc{Ms}. Row 1: original images; row 2-7: images generated by \textsc{Ms}.}
    \label{fig:model-level-protection}
\end{figure}

\begin{figure}[!t]
    \centering
    \includegraphics[width=85mm]{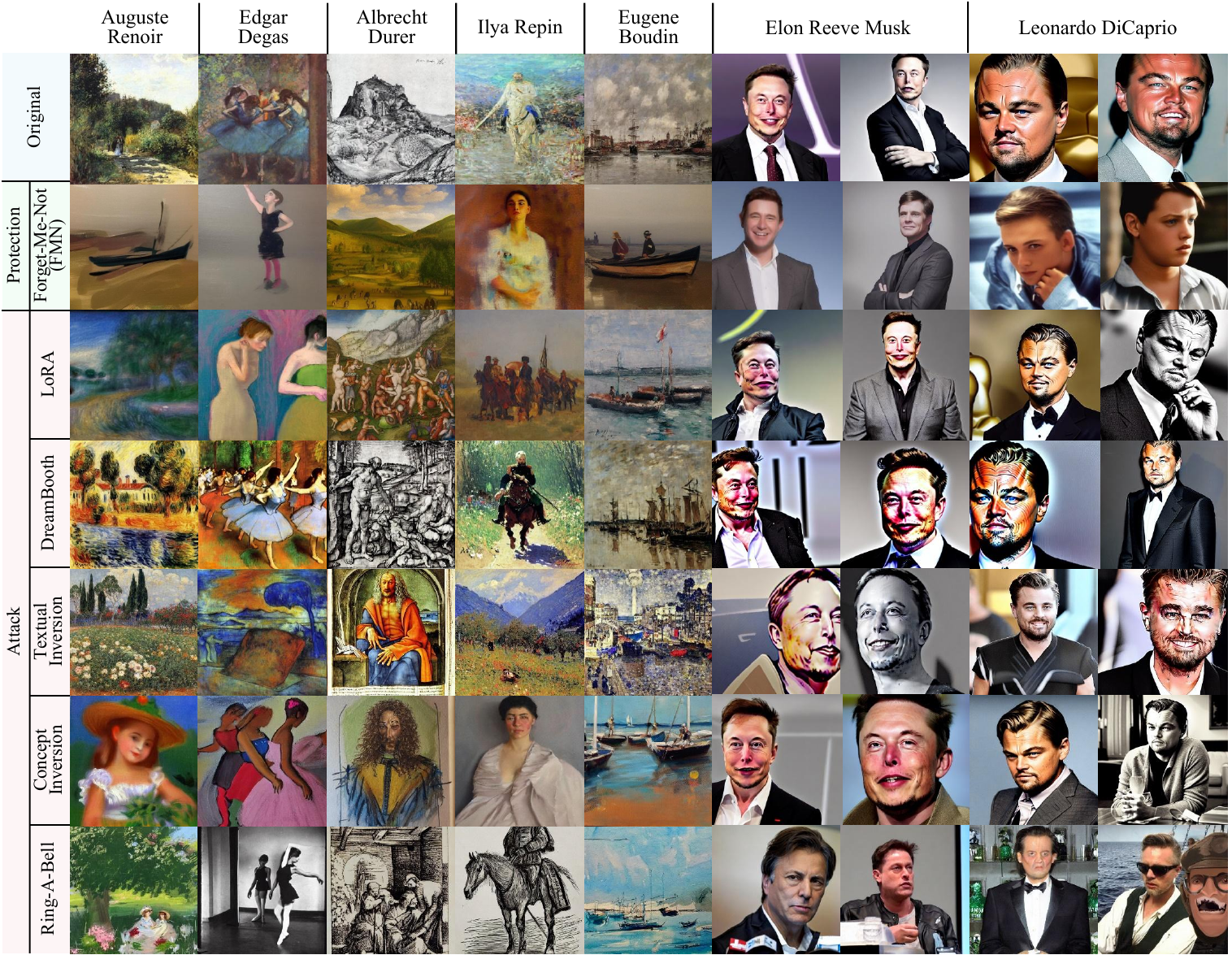}
    \caption{Resilience visualization of \textsc{Ms} against \textsc{Cr}. Row 1: original images; row 2: FMN-sanitized images; row 3-7:  images generated from recovered model.}
    \label{fig:model-level-attack}
\end{figure}

\begin{figure}[ht]
    \centering
    \includegraphics[width=85mm]{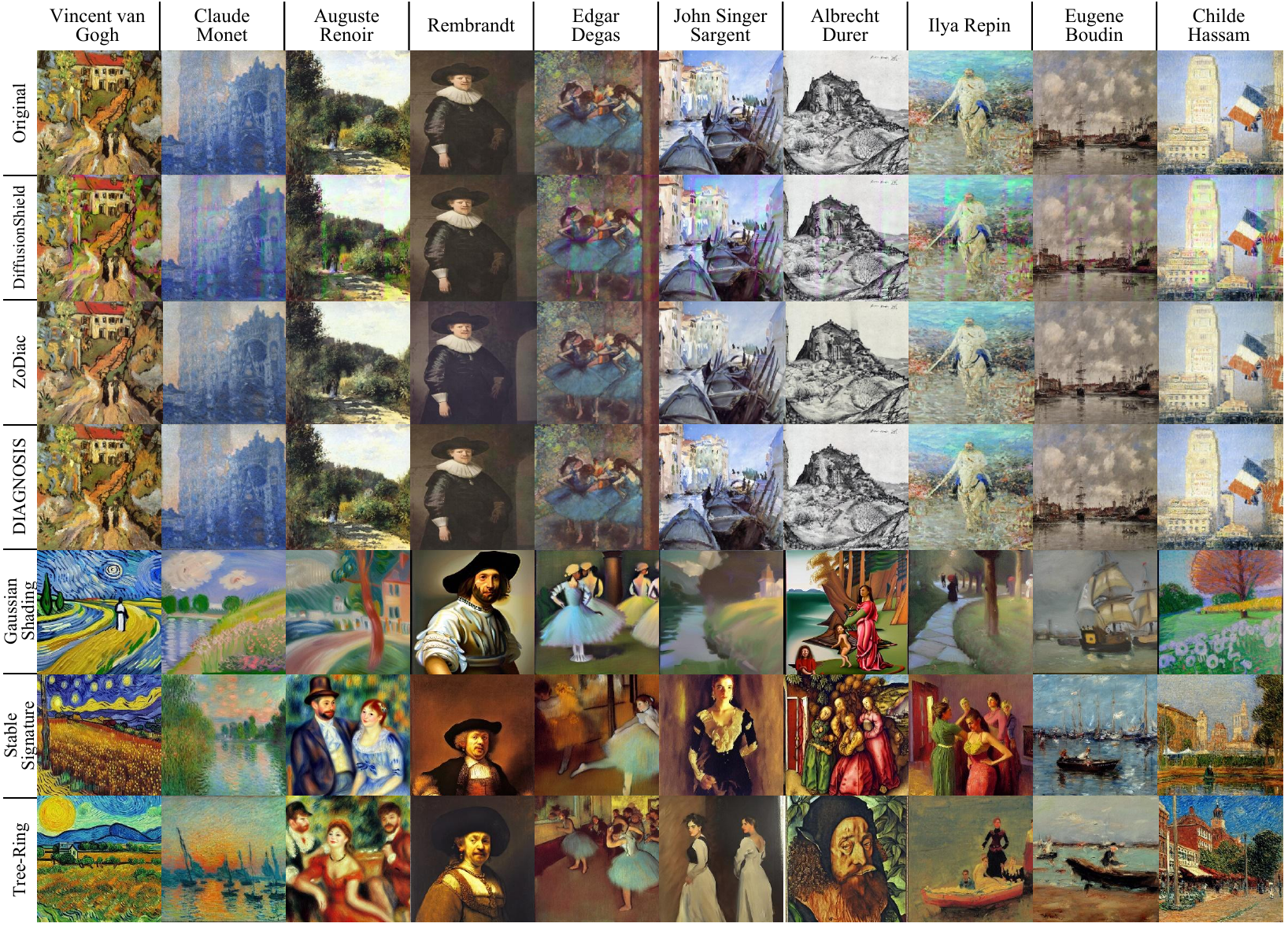}
    \caption{Fidelity visualization of \textsc{Dw}. Row 1: original artworks from WikiArt dataset; row 2-7: watermarked images.}
    \label{fig:output-level-protection}
\end{figure}

\begin{figure}[!t]
    \centering
    \includegraphics[scale=0.36]{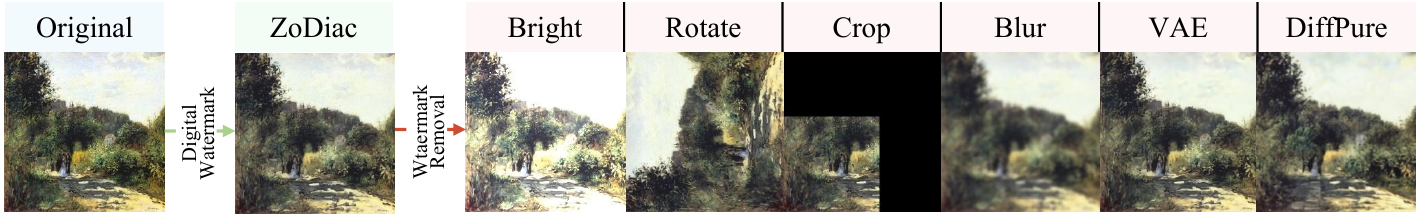}
    \caption{Fidelity visualization of \textsc{Dw} against \textsc{Wr}. Column 1: original image; column 2: Zodiac-watermarked image; column 3-8: watermarked images under attacks.}
    \label{fig:vis-output-att}
\end{figure}
\section{Metrics Overview and Visualization Results}
\label{app:image-result}

\begin{table*}[!t]
{\footnotesize
    \centering
    \renewcommand{\arraystretch}{1}
    \caption{Properties of copyright protection methods.} 
    \setlength{\tabcolsep}{5pt}
    \resizebox{1.0\textwidth}{!}{
    \begin{tabular}{c|M{1.1cm}|p{8cm}|c|c|c|c|c|c|c|c}

    \bf Category & \bf Property& \bf \centering Description & \bf \scriptsize PSNR & \bf \scriptsize SSIM & \bf \scriptsize FID & \bf \scriptsize VIFp & \bf \scriptsize LPIPS & \bf \scriptsize CLIP-I & \bf \scriptsize CLIP-T & \bf \scriptsize ACC \\
    \hline
    \hline
    \multirow{3}{*}{\minitab[c]{Obfuscation\\Processing}}
    & \centering \multirow{1}{*}{Fidelity} & Protected images resemble the originals under all scenarios. & $\checkmark$ & $\checkmark$ & $\checkmark$ & $\checkmark$ & $\checkmark$ & $\times$ & $\times$ & $\times$ \\

    & \centering \multirow{1}{*}{Efficacy} & Protected images mitigate copyright infringement. & $\times$ & $\times$ & $\checkmark$ & $\times$ & $\times$ & $\checkmark$ & $\checkmark$ & $\times$ \\

    & \centering \multirow{1}{*}{Resilience} & Protected images mitigate copyright mimicking under attack. & $\times$ & $\times$ & $\checkmark$ & $\times$ & $\times$ & $\checkmark$ & $\checkmark$ & $\times$ \\
    \hline
    \multirow{3}{*}{\minitab[c]{Model\\Sanitization}}
    & \centering \multirow{1}{*}{Fidelity} & Sanitized models are unaffected for unrelated concepts. & $\times$ & $\times$ & $\checkmark$ & $\times$ & $\times$ & $\times$ & $\checkmark$ & $\times$ \\ 

    & \centering \multirow{1}{*}{Efficacy} & Sanitized models forget specific copyright concepts. & $\times$ & $\times$ & $\checkmark$ & $\times$ & $\times$ & $\times$ & $\checkmark$ & $\times$ \\

    & \centering \multirow{1}{*}{Resilience} & Sanitized models struggle to relearn copyright concepts under attack. & $\times$ & $\times$ & $\checkmark$ & $\times$ & $\times$ & $\times$ & $\checkmark$ & $\times$ \\
    \hline
    \multirow{3}{*}{\minitab[c]{Digital\\Watermark}}
    & \centering \multirow{1}{*}{Fidelity} & Watermarked images resemble the originals under all scenarios. & $\checkmark$ & $\checkmark$ & $\checkmark$ & $\checkmark$ & $\checkmark$ & $\checkmark$ & $\checkmark$ & $\times$ \\

    & \centering \multirow{1}{*}{Efficacy} & Watermark extractable from protected images. & $\times$ & $\times$ & $\times$ & $\times$ & $\times$ & $\times$ & $\times$ & $\checkmark$ \\

    & \centering \multirow{1}{*}{Resilience} & Watermark extractable from attacked images. & $\times$ & $\times$ & $\times$ & $\times$ & $\times$ & $\times$ & $\times$ & $\checkmark$ \\
    \end{tabular}
    }
    \label{tab:watermarking_requirements}
}
\end{table*}

We use metrics to assess \textit{fidelity}, \textit{efficacy}, and \textit{resilience} of copyright protection methods, with Table \ref{tab:watermarking_requirements} summarizing these properties across different categories. 
For obfuscation processing and noise purification, Figure \ref{fig:data-level-protection} presents protected images alongside the original artwork, while Figure \ref{fig:data-level-dreambooth-generation} shows DreamBooth fine-tuned images with reference to the protected and purified images. 
For model sanitization and concept recovery, Figure \ref{fig:model-level-protection} shows protected images with a specific concept sanitized, and Figure \ref{fig:model-level-attack} shows the recovered images.
Finally, for digital watermark and watermark removal, the results are illustrated in Figure \ref{fig:output-level-protection}.
\section{Experimental Setup}
\label{app:exp-setup}

\subsection{Obfuscation Processing and Noise Purification}
\label{app:protection-data}
In \textsc{Op}, \textbf{AdvDM} \cite{liang2023adversarial} trains with learning rate of 0.003 for 100 steps, and a perturbation limit of 0.06. \textbf{Mist} \cite{liang2023mist} uses an \(l_{\infty}\) constraint, 100 PGD steps, a per-step perturbation of 1/255, and a total budget of 16/255. Given that \textbf{Glaze} \cite{shan2023glaze} is closed-source, we follow the implementation from \cite{cao2024impress}'s code using a learning rate of 0.001 for 500 steps, with a perceptual perturbation budget of 0.05, LPIPS loss weight of 0.1. \textbf{PhotoGuard (PGuard)} \cite{salman2023raising} uses an \(\ell_{\infty}\) perturbation limit of 16/255, step size of 2/255, and 200 optimization steps. \textbf{Anti-Dreambooth (AntiDB)} \cite{van2023anti} employs 100 PGD iterations for FSMG and 50 for ASPL, with a perturbation budget of 8/255, a step size of 1/255, and a noise budget \( \eta \) of 0.05, minimized over 1000 training steps.

In \textsc{Np}, \textbf{JPEG Compression (JPEG)} \cite{wallace1991jpeg} sets the quality to 0.75, and \textbf{Quantize (Quant)} \cite{heckbert1982color} sets the bit depth to 8. \textbf{Total Variance Minimization (TVM)} \cite{chambolle2004algorithm} sets a regularization weight of 0.5 with the \(l_{2}\) norm and optimized with the BFGS algorithm. For \textbf{IMPRESS} \cite{cao2024impress}, we use the original authors' hyperparameters, setting the learning rate to 0.001, purification intensity to 0.1, and 3000 iterations. For \textbf{DiffPure} \cite{nie2022diffusion}, we use classifier-free guidance with a scale of 7.5 and fine-tune the diffusion timesteps with a strength of 1,000 via AutoPipelineForImage2Image\footnote{\url{https://huggingface.co/docs/diffusers/api/pipelines/auto_pipeline}}.

\subsection{Model Sanitization and Concept Recovery}
\label{app:protection-model}

We use several methods for \textsc{Ms}. For \textbf{Forget-Me-Not (FMN)} \cite{zhang2024forget}, we fine-tune by textual inversion scripts provided by the authors. For \textbf{Erased Stable Diffusion (ESD)} \cite{gandikota2023erasing}, models are trained for each category. \textbf{Ablating Concepts (AC)} \cite{kumari2023ablating} employs scripts from the authors for both artistic and personal concepts, utilizing WikiArt artworks and generated photos, respectively. \textbf{Unified Concept Editing (UCE)} \cite{gandikota2024unified} models are trained using default parameters. \textbf{Negative Prompt (NP)} is applied during inference by txt2img.py\footnote{\url{https://github.com/CompVis/stable-diffusion/blob/main/scripts/txt2img.py}}. \textbf{Safe Latent Diffusion (SLD)} \cite{schramowski2023safe} use a new SD pipeline based on the diffusers\footnote{\url{https://github.com/huggingface/diffusers/}}, with safety concepts defined for both artistic and personal elements. All parameters are set to default: guidance scale at 2000, warm-up steps at 7, threshold at 0.025, momentum scale at 0.5, and momentum beta at 0.7.
Additionally, to confirm sanitized models' fidelity on unrelated concepts, we compare 30,000 real-word images from MS-COCO 2017 dataset and the SD-generated images from the corresponding text descriptions. 

In \textsc{Cr}, \textbf{LoRA} \cite{hu2021lora} trains with a batch size of 1 and a learning rate of \(1 \times 10^{-4}\) for 100 steps. \textbf{DreamBooth (DB)} \cite{ruiz2023dreambooth} trains with a batch size of 2 and a learning rate of \(5 \times 10^{-7}\) for 1000 steps, using prompts such as ``\textit{a painting in the style of [V]}'' for WikiArt dataset and ``\textit{A photo of sks [V]}'' for Person dataset, where ``\textit{[V]}'' represents a artist or concept name. \textbf{Textual Inversion (TI)} \cite{gal2022image} uses textual\_inversion.py\footnote{\url{https://github.com/huggingface/diffusers/blob/main/examples/textual_inversion/textual_inversion.py}} with 1000 steps and a learning rate of \(5 \times 10^{-4}\), using the same prompts as DreamBooth. \textbf{Concept Inversion (CI)} \cite{pham2023circumventing} trains with a batch size of 4, a learning rate of \(5 \times 10^{-3}\), and 1000 steps, using frozen erased model weights. \textbf{Ring-A-Bell (RB)} \cite{tsai2023ring} uses a prompt length of 77, a tuning coefficient of 3, and a genetic algorithm with a population of 200, 3000 iterations, a mutation rate of 0.25, and a crossover rate of 0.5. 
UCE uniquely generates non-standard .pt model files, preventing further fine-tuning. Inference-guiding protections (NP, SLD) do not generate model files and are vulnerable only to CI and RB attacks. 
\subsection{Digital Watermark and Watermark Removal}
\label{app:protection-output}

In \textsc{Dw}, \textbf{DiffusionShield (DShield)} \cite{cuidiffusionshield} uses a patch shape of \((u, v) = (4, 4)\) and sets a quarternary message to 2. For joint optimization, a 5-step PGD \cite{madry2017towards} is applied with \(l_\infty \leq \epsilon\), while SGD optimizes the classifier. \textbf{Diagnosis (Diag)} \cite{wang2023diagnosis} uses a 100\% coating rate for unconditional and 20\% for trigger-conditioned memorization, with warping strengths of 2.0 and 1.0, respectively. \textbf{Stable Signature (StabSig)} \cite{fernandez2023stable} fine-tunes the LDM decoder using decoder to generate watermarked images. \textbf{Tree-Ring (TR)} \cite{wen2023tree} uses guidance scale of 7.5 for 50 inference steps, with a watermark radius of 10 for DDIM inversion. \textbf{ZoDiac} \cite{zhang2024robust} uses a pre-trained SD model with 50 denoising steps and optimizes the latent variable over 100 iterations, using a watermark radius of 10 and weights of 0.1 for SSIM loss and 0.01 for perceptual loss. \textbf{Gaussian Shading (GShade)} \cite{yang2024gaussian} samples 50 steps using DPMSolver \cite{lu2022dpm} with a guidance scale of 7.5 and performs 50 steps of DDIM inversion, using settings of \(f_c = 1\), \(f_{hw} = 8\), and \(l = 1\) with capacity of 256 bits. 

In \textsc{Wr}, \textbf{Brightness Adjustment (Bright)} \cite{yang2024gaussian} applies a factor of 6, and \textbf{Image Rotation (Rotate)} \cite{yang2024gaussian} performs a 90-degree rotation. \textbf{Random Crop (Crop)} \cite{yang2024gaussian} executes a selection of 50\% of the image area, while \textbf{Gaussian Blur (Blur)} \cite{hosam2019attacking} uses a kernel size of 4. \textbf{VAE-Cheng20 (VAE)} is utilized with a quality level of 3 \cite{cheng2020learned}. Moreover, \textbf{DiffPure} \cite{nie2022diffusion} implements the AutoPipelineForImage2Image pipeline\footnote{\url{https://huggingface.co/docs/diffusers/api/pipelines/auto_pipeline}}, with classifier-free guidance set to 7.5 and diffusion timesteps tuned to a strength of 1,000.



\subsection{Style Mimicry Experimental Details}
\label{app:minicry-dreambooth}
Dreambooth is a subject-driven generation method that can be used for style/concept transfer. In \textsc{Op} and \textsc{Np}, we use unprotected, protected, and attacked images as references to fine-tune a pre-trained SD model via Dreambooth, utilizing the implementation provided by diffusers\footnote{\url{https://github.com/huggingface/diffusers/}}. 
\minxi{Additionally, we use the T2I fine-tuning script provided by diffusers to test the generalization of the protections (\textit{cf}. Section \ref{sec:generalizability}). Following \cite{honig2024adversarial} for optimal style mimicry, we use 2000 training steps, a batch size of 4, a learning rate of \(5 \times 10^{-6}\).} 

\section{Implementation \tbd{Validity} Analysis of Glaze}
\label{app:correctness-analysis}

\begin{figure}[!t]
    \centering
    \includegraphics[width=85mm]{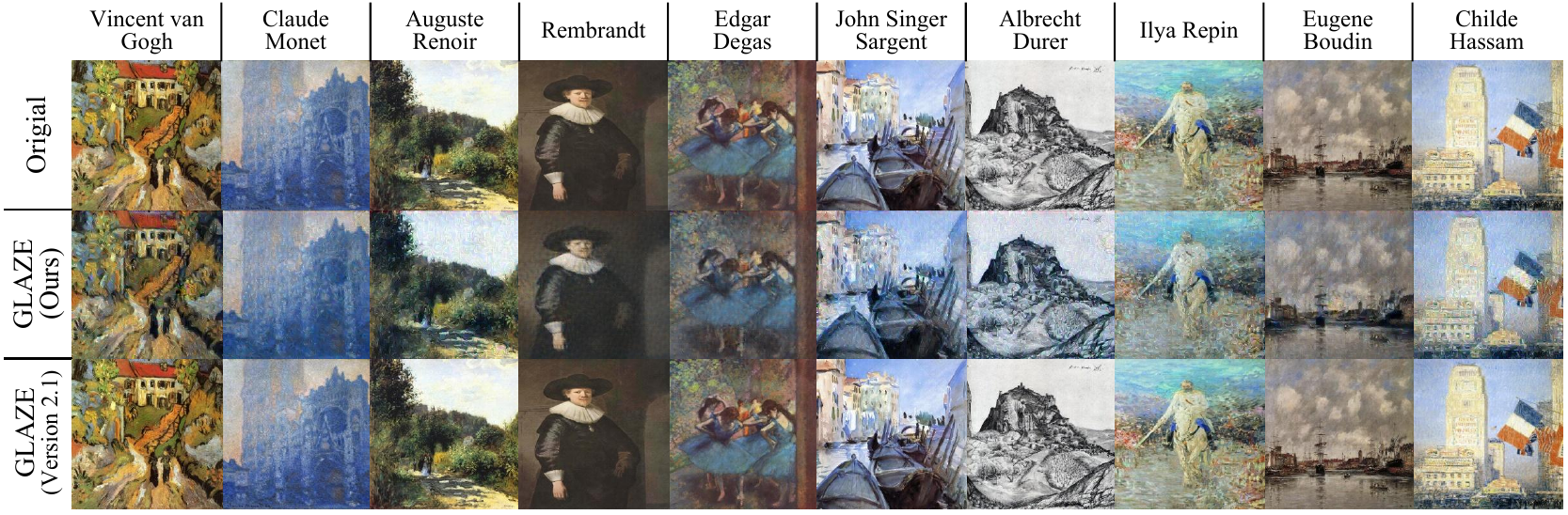}
    \caption{The comparison of generated images of a simplified version of the Glaze with Glaze v2.1.}
    \label{fig:comparision-of-Glaze}
\end{figure}

We use Glaze's reproduce code from IMPRESS\footnote{\url{https://github.com/AAAAAAsuka/Impress/blob/main/glaze.py}} since the latest version of Glaze (v2.1)\footnote{\url{https://Glaze.cs.uchicago.edu/downloads.html}} is not open-sourced. For Glaze v2.1, we set the intensity to high and render quality to slowest for maximum protection. 
The comparison of protected images shows that while our implementation offers slightly lower protection, it achieves higher fidelity (\textit{cf}. Table \ref{tab:comparison-on-Glaze}). Both approaches display similar ``style cloaks,'' confirming the validity of our implementation (\textit{cf}. \ref{fig:comparision-of-Glaze}).


\begin{table}[!t]
\centering
{\footnotesize
    \centering
    \renewcommand{\arraystretch}{1}
    \caption{Comparison of fidelity and efficacy on Glaze.}
    \setlength{\tabcolsep}{2pt}
    \resizebox{1.0\linewidth}{!}{
    \begin{tabular}{c|c|c|c|c}
    {\bf Method} & {\bf LPIPS $\downarrow$} & {\bf FID $\downarrow$} & {\bf CLIP-I $\downarrow$} & {\bf CLIP-T $\downarrow$} \\
    \hline
    \hline
    Glaze v2.1 & 0.403 $\pm$ 0.053 & 283 & 0.625 $\pm$ 0.008 & 0.248 $\pm$ 0.002 \\
    Our Implementation & 0.133 $\pm$ 0.031 & 182 & 0.698 $\pm$ 0.010 & 0.292 $\pm$ 0.001 \\
    \end{tabular}
    }
    \label{tab:comparison-on-Glaze}
}
\end{table}

\section{User Study}
\label{app:appendix-user-study}


\textbf{User Study of \textsc{Op}.}
Our human evaluation assesses both visual quality and style mimicry of protected images under various attacks. Following \cite{shan2023glaze,honig2024adversarial}, we measure the correlation between metrics and human judgment regarding artist style mimicry. Annotators on Amazon MTurk\footnote{\url{https://www.mturk.com/}} were presented with original artworks as style references and asked to evaluate two scenarios: (\textit{i}) a generated artwork without protection versus one with protection, and (\textit{ii}) a generated artwork without protection versus one with protection after attack.
We employ original artist images from the WikiArt and the corresponding protected images from different protection methods as reference pictures to fine-tune the Dreambooth model with a prompt ``a painting in the style of [\textit{artist}]''. 
Participants view 10 original artworks by a specific artist as reference samples, followed by one protected and one unprotected generated image in the same style. We focus on two key aspects: 
1) \textit{Visual Quality.} Participants assess each image based on four questions corresponding to metrics targeting noise level, fidelity (including artifacts), alignment with brightness/contrast/structure, and overall stylistic fit (\textit{cf}. Table \ref{tab:question_list}). To ensure unbiased assessments, we randomized image order, comparison sequences, and model generation seeds. 
2) \textit{Style Mimicry.}
Inspired by the Glaze \cite{shan2023glaze}, we asked participants to rate the style mimicry of the generated images on a 5-point Likert scale, evaluating how well each image resembled the reference style samples. 
The options range from: (i) \textit{Not successful at all}, (ii) \textit{Not very successful}, (iii) \textit{Somewhat successful}, (iv) \textit{Successful}, to (v) \textit{Very successful}.

\begin{figure}[t]
    \includegraphics[width=\linewidth]{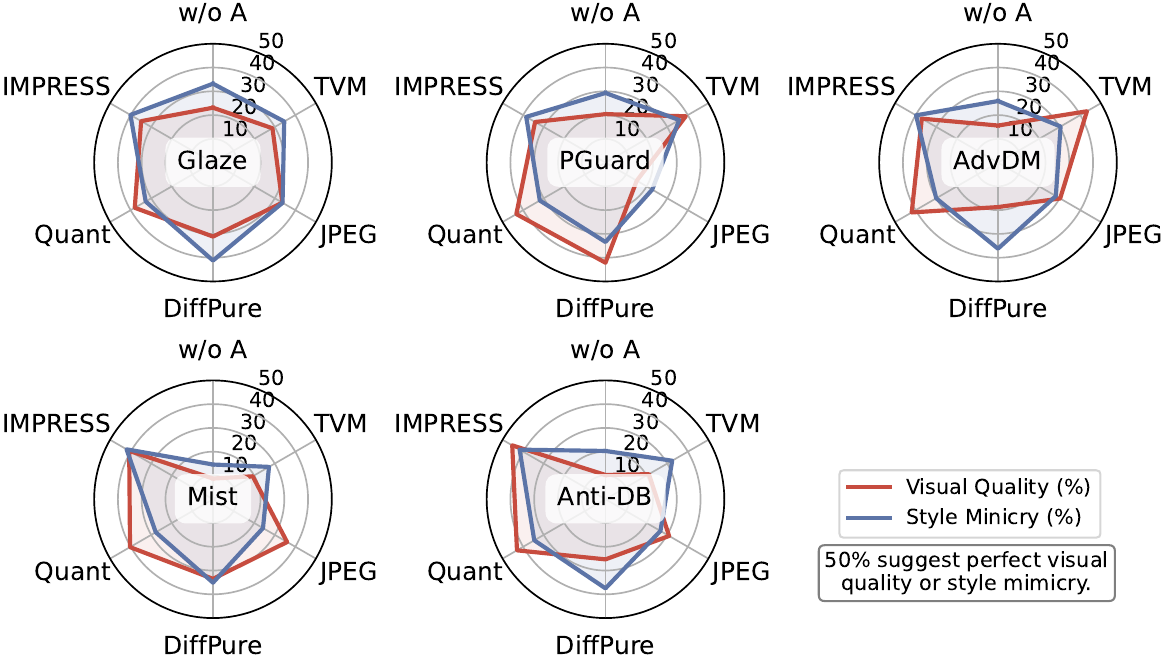}
     \centering
    \caption{Visual quality and style mimicry success rates across different \textsc{Op} protection against attacks.}
    \label{fig:success-rate}
\end{figure}

\begin{table}[!t]
{\footnotesize
    \centering
    \renewcommand{\arraystretch}{1}
    \caption{Question list for \textsc{Op}.}
    \setlength{\tabcolsep}{2pt}
    \resizebox{\linewidth}{!}{
    \begin{tabular}{c|c|p{8.4cm}}
    \multicolumn{2}{c|}{\textbf{Dimensions}} & \textbf{Human Evaluation Questions} \\
    \hline
    \hline
    \multirow{6}{*}{\minitab[c]{Visual\\Quality}} & PSNR & Which image has less noise? \\
    \cline{2-3}
    & VIFp & Which image has better fidelity and fewer artifacts (distorted, unrealistic)? \\ 
    \cline{2-3}
    & SSIM & \multirow{2}{8.4cm}{Based on brightness, contrast, and structure, which better matches the\\ referred image?} \\
    \cline{2-2}
    & LPIPS & \\
    \cline{2-3}
    
    & FID & \multirow{2}{8.4cm}{Which image better fits the style of the referred image samples and the \\description ``a painting in the style of [\textit{artist}]''?}\\
    \cline{2-2}
    & CLIP-I/T & \\
    \hline
    \multicolumn{2}{c|}{\multirow{1}{*}{\minitab[c]{Style Mimicry}}}  &  How successfully does the style of the image mimic the samples? \\
    
    \end{tabular}
    }
    
    \label{tab:question_list}
}
\end{table}






\textbf{Fidelity of \textsc{Ms}. }
We conduct this user study to explore whether the \textsc{Ms} methods would impact the fidelity of images of unrelated concepts from human's perspective.
We evaluate image fidelity and text alignment by generating 2000 images per \textsc{Ms} model. Participants assess 25 random image pairs, comparing the SD reference to an erased model image, answering two questions: (i) \textit{Which image is of higher quality?} (ii)\textit{ Which image better represents the text caption?} For each pair, the participants could respond with: (i) \textit{I prefer image A}, (ii) \textit{I am indifferent}, or (iii) \textit{I prefer image B}. The study is conducted via Amazon Mechanical Turk, requiring participants to have a HIT Approval Rate above 95\% and at least 1000 approved HITs. Each image pair batch is evaluated by three annotators, with each prompt receiving 30 assessments. 

\begin{table}[t]
{\footnotesize
    \centering
    \caption{Image fidelity and text alignment of \textsc{Ms}.}
    \renewcommand{\arraystretch}{1.0}
    \setlength{\tabcolsep}{2pt}
    \begin{tabular}{c|c|c||c|c|c|c|c|c}
    & \textbf{Method} & \textbf{SD} & \textbf{FMN} & \textbf{ESD} & \textbf{AC} & \textbf{UCE} & \textbf{NP} & \textbf{SLD} \\ 
    \hline
    \hline
    {\bf \multirow{2}{*}{\minitab[c]{Image\\Fidelity}}} & \textbf{FID-30k} $\downarrow$ & \textbf{16.21} & 16.47 & 16.51 & 16.95 & 16.64 & 16.89 & 16.95 \\ 
    \cline{2-9}
    & \textbf{User/\%} $\uparrow$ & - & 62.93 & 63.02 & 63.87 & 63.56 & 63.50 & 63.98\\ 
    \hline
    {\bf \multirow{2}{*}{\minitab[c]{Text\\Alignment}}} & \textbf{CLIP-T} $\uparrow$ & \textbf{0.31} & 0.30 & 0.30 & 0.31 & 0.31 & 0.30 & 0.30 \\ 
    \cline{2-9}
    &  \textbf{User/\%} $\uparrow$ & - & 59.37 & 59.46 & 61.04 & 60.89 & 59.51 & 59.38  \\
    \end{tabular}
    \label{tab:user-ms}
}
\end{table}

\section{Extentions to Other Image Editing Tasks}
Our experiments with methods like AdvDM, Mist, and Glaze focused on unauthorized style imitation. It is essential to assess whether these protections also prevent unauthorized attribute editing. PGuard \cite{salman2023raising}, originally designed for style imitation, applies imperceptible adversarial perturbations to handle a range of unauthorized edits, indicating potential for broader editing protection. For PGuard, we use Img2ImgPipeline\footnote{\url{https://huggingface.co/docs/diffusers/api/pipelines/stable_diffusion/img2img}} for image editing. For the WikiArt dataset, we guide the SD model with style transformation prompts, such as: ``Transform Vincent van Gogh's `Starry Night' into a surrealist painting in the style of Salvador Dalí.'' For the CustomConcept101 dataset, we use prompts like ``Change the background to a snowy mountain landscape during sunset while keeping the person unchanged.'' 

\begin{figure}[ht]
    \includegraphics[width=85mm]{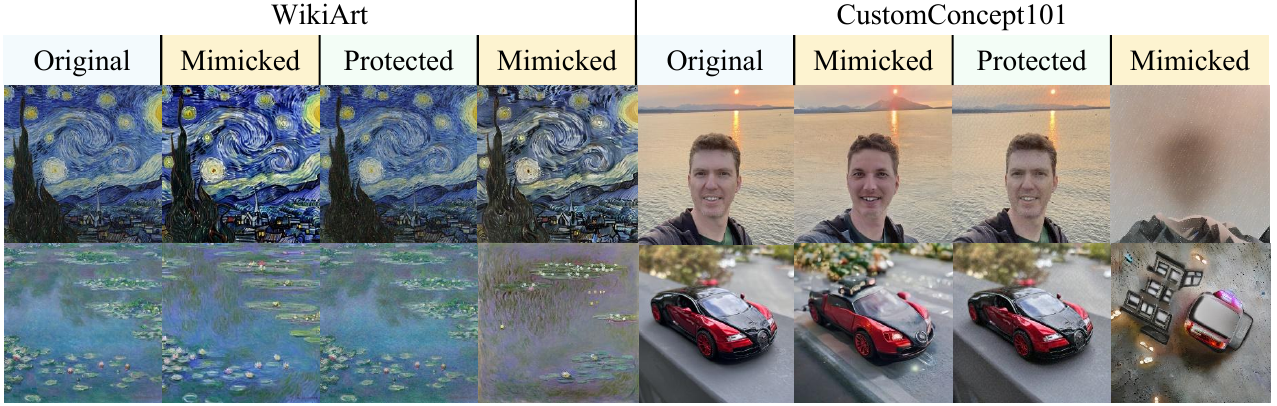}
     \centering
    \caption{The result of PGuard protection.}
    \label{fig:pguard1}
\end{figure}

\begin{figure}[ht]
    \includegraphics[width=85mm]{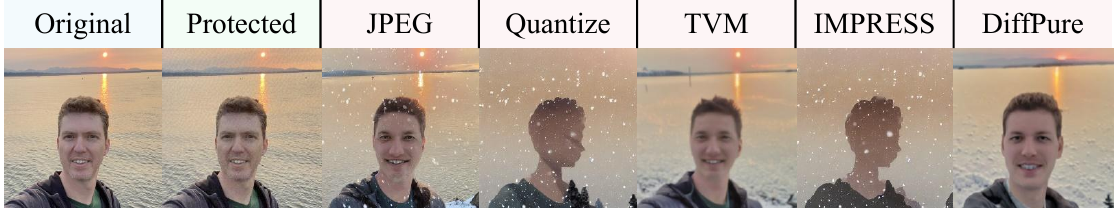}
     \centering
    \caption{The result of PGuard’s protection under attacks.}
    \label{fig:pguard2}
\end{figure}

In examining PGuard’s resilience against various \textsc{Np} methods, Figures \ref{fig:pguard1} and \ref{fig:pguard2} reveal that while PGuard effectively prevents unauthorized edits, it faces challenges under specific attack scenarios. These attacks can lead to I2I transformations and result in images that merge original features with prompt modifications. This aligns with Section \ref{sec:op-evaluation}, highlighting that no \textsc{Op} protection is entirely resistant to \textsc{Np} attacks. Our findings suggest that protection effectiveness is closely linked to the intended application context, underscoring the need for further exploration into broader challenges like attribute editing.


\end{document}